\documentclass{article}

\usepackage{graphicx}
\usepackage{amsmath}
\usepackage{amssymb}
\usepackage{amsthm}
\usepackage{mathtools}
\usepackage{hyperref}
\usepackage{booktabs}
\usepackage{xcolor}
\usepackage{siunitx}
\usepackage[normalem]{ulem}
\usepackage[caption=false]{subfig}
\usepackage[numbers, compress, square]{natbib}
\usepackage{fancyhdr}
\usepackage[margin=1.5cm]{geometry}
\usepackage{authblk}

\newcommand{\wasserstein}{\mathcal{W}}

\DeclareMathOperator*{\argmin}{argmin}
\DeclareMathOperator{\conv}{conv}
\DeclareMathOperator{\Rips}{Rips}

\theoremstyle{definition}
\newtheorem*{definition}{Definition}

\begin{document}
    
\title{Fitting the topology of synthetic particle systems \\ with a novel graph representation}
\author{Martin Alexander Memmesheimer}
\author{Claudia Redenbach}

\affil{\small martin.memmesheimer@rptu.de, claudia.redenbach@rptu.de}
\affil{\small \begin{tabular}{c}
Department of Mathematics, \\ RPTU University Kaiserslautern-Landau, \\ Gottlieb-Daimler Str.\ 30, 67663 Kaiserslautern, Germany \end{tabular}}

\date{}

\maketitle
\thispagestyle{fancy}

\begin{abstract}
    The shape and arrangement of particles in a material determine its macroscopic properties. The generation of synthetic data with varying particle structure, often represented as 3D voxel images, combined with simulation of macroscopic properties reveals structure-property relations. Most particle generation models focus on single-particle characteristics like shape and size. We aim at fitting the topology of the particle system using tools from persistent homology. However, the large size of the required 3D image data makes existing methods computationally infeasible. We bridge this gap by introducing a novel graph representation of particle systems and transferring the computation of persistent homology from the image domain to the graph domain. This yields a postprocessing method for synthetic images of particle systems, that is independent of the underlying generation method and improves topological and geometrical agreement with real particle systems while preserving morphological characteristics such as the particle size distribution.
\end{abstract}

\section{Introduction}
\label{Section: Introduciton}

Particle systems occur in a wide range of natural and engineered materials, including sand, gravel, microdiorite, concrete, iron ore or pigments in coatings. Their mechanical behavior has been the subject of extensive research~\cite{Farhang_al2017GranularMaterialsDecade}. Both the geometry of individual particles~\cite{Erdogan_al2006EngineeringApplication,Fonseca_al2012EngineeringApplication, Garboczi2002EngineeringApplication} and the topology and connectivity of the particle system~\cite{Liu_al2020ConnectivityEngineeringApplication,Papadopoulos_al2018TopologyEngineeringApplication, Walker_Tordesillas2010TopologyEngineeringApplication} have been shown to strongly influence its macroscopic behavior~\cite{Lin_al2022MacroscopicBehaviour, ZhaoandZhou2017MacroscopicBehaviour}. In other applications, such as active protective coatings, leaching of pigment particles is the determining process~\cite{ZANINOVIC2026113631}. Also in this case, the connectivity of the particle system is decisive for the coating performance.

X-ray micro-computed tomography ($\mu$CT) provides the capability to capture the complex three-dimensional geometry of particle systems. However, the acquisition of $\mu$CT images requires specialized and costly equipment. As a result, only limited amounts of data are available. Thus, the use of data-driven approaches for the analysis of microstructure-property relationships in particulate materials is limited.

The generation of synthetic data is a promising way to circumvent the issues posed by a lack of data. A key component of the generation pipeline is the placement of particles, for which various strategies have been proposed, including random sequential adsorption~\cite{chiu_al2013stochasticgeometry}, matching morphological descriptors~\cite{You_al2019SyntheticParticleComposites}, random tessellations~\cite{Feinhauer_al2015ParticleModellingSphericalHarmonics}, and rigid body simulations~\cite{Wang_al2025SyntheticParticlePack}, which aim to reproduce various statistical and physical characteristics of particle systems. Many downstream applications require an accurate reproduction of the topological structure, including connectivity and higher-dimensional topological features, which existing generation methods typically do not explicitly optimize.

In recent years, there has been a growing interest in integrating topological data analysis, in particular persistent homology, into synthetic data generation pipelines~\cite{Gupta_al2025TopoDiffusionNet,Liu_al2025TopoLiDM, Wang_al2020TopoGAN}. Persistent homology is a powerful tool that captures the topological characteristics of data through a summary statistic known as the persistence diagram. By incorporating persistence homology based loss terms, topological optimization has also been applied to a variety of tasks outside of synthetic data generation, including surface matching~\cite{Poulenard_al2018SurfaceMatching}, point cloud processing~\cite{BruelGabrielson_al2020SurfaceReconstructionForPointClouds}, and topological simplification~\cite{Kissi_al2025TopologicalSimplification}. While many existing approaches are tailored to specific applications, Carriere et al.~\cite{Carriere_al2021OptimizingPHbasedFunctions} introduced a general framework for topological optimization. However, the direct application of existing methods to three-dimensional $\mu$CT images of a particle system is computationally prohibitive due to their large size. Furthermore, particle-specific information of the $\mu$CT image cannot be exploited when computing persistent homology in the image domain.

\begin{figure*}
    \centering
    \includegraphics[width=0.75\linewidth]{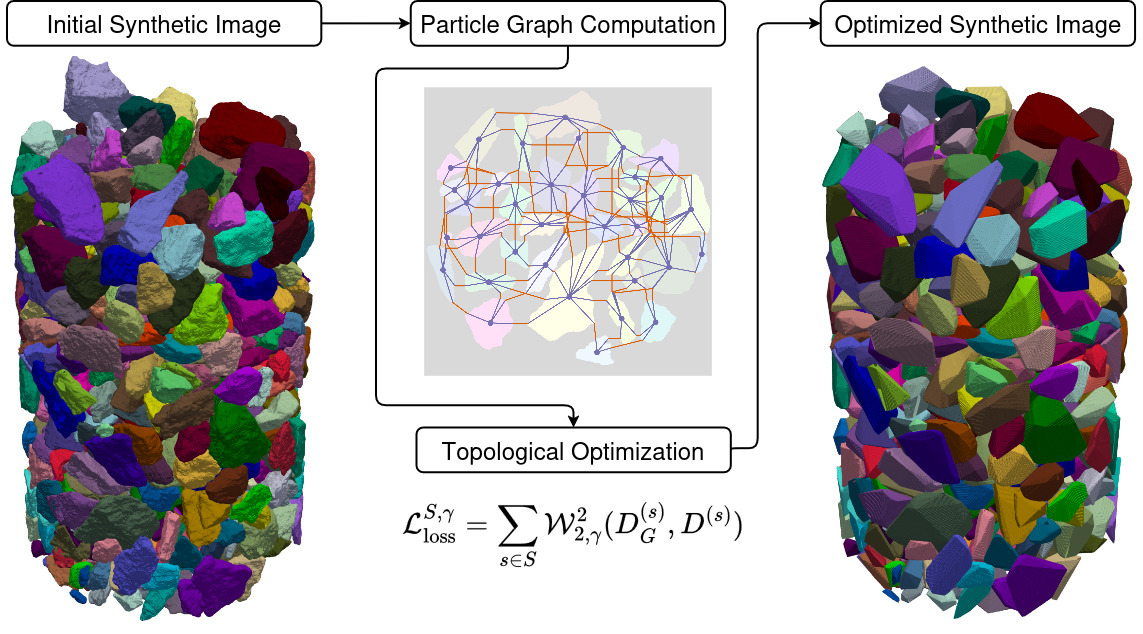}
    \caption{Graphical overview of our proposed method.}
    \label{Figure: Graphical abstract}
\end{figure*}

To address these issues, we propose a novel graph representation of particle systems, which enables topological optimization as a postprocessing step for synthetic images, independent of the underlying generation method. While graph representations of particles already exist, they usually do not contain enough information to transition back to the image domain. Our graph representation is specifically designed to contain the necessary information to allow this reconstruction, while also providing a meaningful representation of the image topology and a fast topological optimization algorithm. Our approach is illustrated in Figure~\ref{Figure: Graphical abstract} and consists of three stages. First, we compute the graph representation on both a real and synthetic image. Next, we apply topological optimization in the graph domain to align the topology of the synthetic graph with that of the real graph. Finally, we return to the image domain by reconstructing an image from the particle graph, whose structure enables an accurate recovery of both particle shapes and the topology of the particle system. Our proposed graph representation offers several advantages for the topological optimization. It leverages particle-specific information while simultaneously reducing the computational cost. Furthermore, it provides a natural interpretation of the optimization in the image domain. In particular, each gradient update in the graph domain can be interpreted as a local deformation of the particle boundary in the image domain. 

The remainder of this paper is structured as follows. In Section~\ref{Section: Background on Persistent homology} we provide a brief introduction to the persistent homology computation on graphs and images, as well as a metric between persistence diagrams to quantify the topological difference in data. In Section~\ref{Section: Graph Representation of Particle Systems} we introduce our graph representation and the reconstruction of the image from the graph, and discuss the relation between the image and graph topology. The topological optimization is explained in Section~\ref{Section: Topological Optimization}. There we also discuss the advantages of optimizing on our graph representation and not directly on the image. Finally, we validate our method on recycled concrete data and show its applicability for leaching in active protective coatings in Section~\ref{Section: Experimental Validation}. Limitations and future research directions are discussed in Section~\ref{Section: Discussion}.

\section{Background on Persistent homology}
\label{Section: Background on Persistent homology}
In this section, we introduce persistent homology and describe its application to both images and graphs. Its foundation was laid in the works of Edelsbrunner et al.~\cite{Edelsbrunner_al2002Intro_PH} and Zomorodian and Carlson~\cite{Zomorodian_Carlson2005Intro_PH} with applications to point clouds. Since then, it has been applied to various types of data, including images and graphs.

Topology studies properties that are invariant under continuous deformations. For example, a circle and an ellipse are topologically equivalent. The topological features of interest for this work are connected components ($0$D), loops ($1$D, e.g. the tunnel in a torus), and cavities ($2$D, e.g. the interior of a sphere). The general idea of persistent homology is to associate data with a topological space and study how its topological features change along a filtration, i.e., a nested sequence of topological subspaces ending in the full space.

To efficiently compute persistent homology, it is important to have combinatorial instead of geometrical descriptors of the topological spaces. For graphs, one typically uses simplicial homology. It can also be used for image data  after triangulating the image domain, e.g., using a Freudenthal triangulation~\cite{Freudenthal1942Triangulation}. Here, we instead use cubical homology, which operates directly on the voxel grid and therefore provides a natural framework for image data. We introduce only the basic ideas and omit many technical details. For a comprehensive introduction to simplicial persistent homology, we refer the reader to Edelsbrunner and Harer~\cite{Edelsbrunner_Harer2010IntroComputationalTopology}, and for cubical homology to Kaczynski et al.~\cite{Kaczynski_al2004IntroCubicalHomology}.

\subsection{Simplicial Homology on Graphs}
\label{Subsections: Simplicial Homology on Graphs}
Simplicial homology is based on (abstract) simplicial complexes. An (abstract) simplicial complex on a vertex set $V$ is a family $K$ of subsets of $V$ that satisfies the following properties:
\begin{enumerate}
    \item For every $v \in V$, $\{v\}$ is in $K$.
    \item For every $\sigma \in K$, every subset of $\sigma$ is also contained in $K$.  
\end{enumerate}
The elements of $K$ are called simplices. The dimension of a simplex $\sigma \in K$ is given by $\vert \sigma \vert - 1$, and the dimension of $K$ is the maximum dimension of its simplices. 

Simplicial complexes are a generalization of (undirected) graphs $G=(V,E)$ with node set $V$ and edge set 
\begin{equation}
    E \subset \{M \subseteq V \ \vert \ \vert M \vert = 2 \}. 
\end{equation}
By identifying a vertex $v \in V$ with the singleton $\{v\}$, a graph can be viewed as a one-dimensional simplicial complex $V \cup E$. Although it is possible to study the topology of a graph as a one-dimensional simplicial complex, it is usually more informative to construct a higher-dimensional simplicial complex from it. 

A widely used construction is the Vietoris-Rips (VR) complex. Given a set of points $\mathbb{X}$, a parameter $\alpha \geq 0$, and symmetric distance function $d \colon \mathbb{X} \times \mathbb{X} \rightarrow \mathbb{R}_\geq$, it is defined by
\begin{equation}
    \sigma \in \Rips_\alpha(\mathbb{X}) \iff \forall x,y \in \sigma \colon d(x, y) \leq \alpha.
\end{equation}
For a graph $G$, whose edges are equipped with a positive weight function $w \colon E \rightarrow \mathbb{R}_\geq$, a distance function can be defined by
\begin{equation}
    \label{Equation: Distance Matrix of Vietoris-Rips}
    d(i,j) = \left\{ \begin{array}{cl} 
        w(\{i,j\}), & \text{if } \{i,j\} \in E \\
        0, & \text{if } i=j \\
        \infty, & \text{otherwise}.
    \end{array} \right.
\end{equation}
The corresponding VR complex is denoted by $\Rips_{\alpha}(G)$. Varying $\alpha$ induces a filtration $(\Rips_{\alpha}(G))_{\alpha \geq 0}$.

\subsection{Cubical Homology on Images}
\label{Subsection: Cubical Homology on Images}

A three-dimensional image is formally represented by a function $\mathcal{I} \colon \Omega \rightarrow \mathbb{R}$, where $\Omega \subset \mathbb{Z}^3$ is the image domain. Depending on the image type, the range of $\mathcal{I}$ is restricted to a subset of $\mathbb{R}$, e.g.\ to grayscale intensities, binary values, or discrete labels. The elements $p \in \Omega$ are called pixels or voxels.

The application of persistent homology via cubical homology to an image is based on an embedding of the discrete image domain $\Omega$ into $\mathbb{R}^3$. Each voxel $\omega = (\omega_1, \omega_2, \omega_3) \in \Omega$  is associated with the cube 
\begin{equation}
    [\omega_1, \omega_1 + 1] \times [\omega_2, \omega_2 + 1] \times [\omega_3, \omega_3 + 1] \subset \mathbb{R}^3.    
\end{equation}
The union of all such cubes is a subset of $\mathbb{R}^3$, which inherits the subspace topology. A filtration is then obtained from the sublevel sets $(f^{-1}((-\infty, r]))_{r \in \mathbb{R}}$ of a function $f \colon \Omega \rightarrow \mathbb{R}$. Each sublevel set is associated with a topological space given by the union of the embedded voxels whose function values do not exceed $r$. 

For grayscale images $\mathcal{I}$, sublevel sets can be computed directly on $\mathcal{I}$. For binary images, this yields limited information, since the sublevel sets change only at two threshold values. Therefore, $f$ is usually obtained by a distance transform 
\begin{equation}
    f(p) = \min_{q \in \Omega} \{\Vert p - q \Vert_1 \ \vert \ \mathcal{I}(q) = 1 \},
\end{equation}
which assigns every pixel $p$ its distance to the nearest foreground pixel. Labeled images are first binarized by setting all pixels $q \in \Omega$ with $\mathcal{I}(q) \neq 0$ to $1$ before computing the distance transform.

\subsection{Persistence Diagrams}
\label{Subsection: Persistence Diagrams}

The classical summary statistic of a persistent homology computation is a collection of persistence diagrams, one for each dimension of topological features. For each dimension $s$, the persistence diagram consists of points $(b,d)$, which lie above the diagonal $\Delta = \{(x, x) \in \mathbb{R}^2 \ \vert \ x \in \mathbb{R} \}$ in $\mathbb{R}^2$. Each point represents an $s$-dimensional topological feature that appears in the filtration at birth time $b$ and disappears at death time $d$. Formally, a persistence diagram is a finite multiset of points supported on 
\begin{equation}
    \mathbb{R}_>^2 = \{(b,d) \in \mathbb{R}^2 \ \vert \ b < d \}.
\end{equation}
In addition, the diagonal $\Delta$ is included with infinite multiplicity. This is required to ensure the existence of a matching between two diagrams, which classical metrics rely on.

We note that some topological features may persist throughout the filtration and never disappear. Such features have infinite death time and are referred to as essential features. In particular, at least one feature in the $0$-dimensional topology is essential, since at least one connected component must persist through the filtration. In higher dimensions, essential features may occur depending on the data. In our application, they never occur in persistence diagrams computed from images and very rarely occur in diagrams computed from graphs. Since essential points are difficult to handle within the optimization framework, we discard them from the persistence diagrams. As their proportion in our data is well below one percent, the information loss due to their omission is negligible. In the subsequent analysis, we always assume that no essential points exist. 

To define a loss function for the optimization framework, we require a distance measure between the persistence diagram of the synthetic and real data. A classical metric is the $2$-Wasserstein distance. Let $D_1, D_2$ be two persistence diagrams with finitely many off-diagonal points, as is always the case for our data. The $2$-Wasserstein distance is defined as
\begin{equation}
    \label{Equation: Definition Wasserstein Distance}
    \wasserstein_2(D_1, D_2) = \left( \inf_{T} \sum_{x \in D_1} \Vert x - T(x) \Vert_2^2 \right)^{\tfrac{1}{2}},
\end{equation}
where the infimum is taken over all possible bijections $T \colon D_1 \rightarrow D_2$. We slightly abuse notation in the above definition by denoting the perpendicular distance of $x$ to the diagonal $\Delta$ by $\Vert x - \Delta \Vert_2$, and by setting $\Vert \Delta - \Delta \Vert_2 = 0$. The inclusion of the diagonal in the persistence diagrams ensures that a bijection exists. Consequently, the feasible set of bijections is nonempty. Since all costs are nonnegative, the objective function in the minimization problem is bounded from below by $0$ such that the Wasserstein distance is well-defined and finite.

\section{Graph Representation of Particle Systems}
\label{Section: Graph Representation of Particle Systems}
\subsection{Graph Construction}
\label{Subsection: Graph Construction}

Applying the topological optimization framework directly to persistence diagrams computed from the image representation of the particle system has several drawbacks, including high computational cost and the need to binarize the image, which converts explicit particle information into implicit information. In the following, we introduce a novel graph representation of particle systems that mitigates the issues posed by the image representation.

Various graph constructions for the analysis of segmented particle systems have been used in the literature. Most notably, contact and force networks were used for the investigation of properties of particle systems such as their rigidity, heat transfer, and acoustic transmission. A thorough survey is given in~\cite{Papadopoulos_al2018TopologyEngineeringApplication}. The idea of contact networks is to represent each particle by a graph node and have an edge between two particles if they are in contact. Force networks extend contact networks by adding a weight function on the edge set, with the weights representing how load is transmitted through the particle system. Persistent homology has been used for the analysis of force networks, e.g.\ by Kram\'ar et al.~\cite{Kramar_al2014ForceNetworksPH}, but with different input information and a different filtration than that we propose later. Contact and force networks are not suitable for our purpose, since a precise reconstruction of an image representation is not possible when only contact positions but no additional spatial information are available. 

Another class of graphs that is used for the analysis of particle systems are Delaunay graphs computed from the Voronoi tessellation of the particle system~\cite{Mei_al2022ProximityGraph, Duran_al2010ProximityGraph}. They contain an edge between two particles whenever their corresponding Voronoi cells share a boundary. The Delaunay graph is a supergraph of the contact graph as it introduces not only edges for contacts between particles, but also for spatially related particles. This is similar to our proposed graph, since we allow for edges if an unobstructed line segment between two particles exists. This also leads to a supergraph of the contact graph, since every contact between two particles yields an unobstructed line segment. Since we want to enable a topology-preserving reconstruction of the particle system, we enrich the graph with additional geometric information on its edges.

In particular, we associate with each edge a start and end point of a shortest unobstructed line segment. The image is then reconstructed by computing the convex hull of all edge features associated with a given particle. We now formalize this construction. 

The image domain of a labeled three-dimensional image $\mathcal{I} \colon \Omega \rightarrow L$ with label set $L = \{0, \ldots, n\}$ can be partitioned into sets $\Omega_i = \{p \in \Omega \ \vert \ \mathcal{I}(p) = i \}$ for $i = 0, \ldots, n$. $\Omega_0$ is called the background set and the union of the individual particles $\Omega_1, \ldots, \Omega_n$ is called the foreground set.

\begin{definition}
    \label{Definition: Particle Graph}
    Given a particle system, represented by a labeled image $\mathcal{I}$. For each pair of foreground labels $u,v$ with $u < v$, we define the set of admissible point pairs
    \begin{equation}
        E_{uv} = \{ (p,q) \in \Omega_u \times \Omega_v \ \vert \  \mathcal{I}(\overline{pq}) \subset \{0,u,v\} \},
    \end{equation}
    where $\overline{pq}$ is the set of voxels on the line segment between $p$ and $q$. The particle graph $G=(V,E,f_V,f_E)$ is then defined by
    \begin{itemize}
        \item  $V = L \setminus \{0\}$.
        \item $E = \{\{u,v\} \ \vert \ E_{uv} \neq \emptyset \}$.
        \item $f_V \colon V \rightarrow \mathbb{R}^3, \ v \mapsto \frac{1}{\vert \Omega_v \vert} \sum_{p \in \Omega_v} p $.
        \item $f_E \colon E \rightarrow \mathbb{R}^6, \{u,v\} \mapsto (p^*,q^*)$ with 
        \begin{equation}
            \label{Equation: argmin_edge}
            (p^{*},q^*) = \argmin_{(p,q) \in E_{uv} }( ||p - q||_1).
        \end{equation}
    \end{itemize} 
\end{definition}

A few remarks on the graph representation: there is a one-to-one correspondence between nodes and particles, and we will use the terms interchangeably in the following. The node features $f_V$ store the centers of mass of the corresponding particles. While this information is not strictly required for the downstream algorithm, it can improve the stability of the reconstruction near the boundaries.

The edge features $f_E$ store the end points of the line segment associated with an edge. These are essential for the reconstruction, as they encode the geometric relationship between two particles. In general, the minimizing argument in Eq.~\eqref{Equation: argmin_edge} is not unique. To ensure that $f_E$ is well defined, we select an arbitrary pair at which the minimum is attained. The condition $u < v$ is included in $E_{uv}$ since it imposes a canonical ordering on the undirected edge $\{u,v\} = \{v,u\}$. 

In practice, we compute the particle graph by selecting the first pair encountered that attains the minimum in Eq.~\eqref{Equation: argmin_edge}. To compute such a pair efficiently, it suffices to consider only pairs of boundary voxels of the respective particles. Admissibility of each such pair $(p,q)$ is determined using Bresenham's line algorithm~\cite{Bresenham1965LineAlgorithm}.

While the full graph contains the most information, the topologically relevant information is expected to be captured primarily by short edges, as features in persistence diagrams are generally associated with spatially localized structures. This motivates using a thresholded version of the graph. For a given threshold $\tau > 0$, the thresholded graph $G_{\tau}$ is obtained by restricting the edge set to 
\begin{equation}
    E_{\tau} = \{e \in E \ \vert \ \Vert p-q \Vert_1 \leq \tau, (p,q) = f_E(e)\}    
\end{equation}
and $f_E$ to $f_E \vert_{E_\tau}$. Computing $G_{\tau}$ is significantly faster than computing the full graph $G$. For simplicity, we denote the thresholded graph by $G$ for the remainder of the paper.

If $\tau$ is chosen appropriately, the persistence diagrams of the thresholded and full graph coincide. In the case of our recycled concrete data (approximate size $800 \times 800 \times 1600$) this is the case for $\tau=150$. A single threaded computation of the thresholded particle graph takes between $20$ and $28$ hours with a peak memory usage of approximately $9.5$ GiB. The variation in computational time is caused by variability of the input images and the memory load on the computing unit due to other processes. While this may appear long, the computation can easily be parallelized. For example, when using $16$ cores, the computation took less than $3$ hours, while the peak memory usage remained almost identical. 

Another way to speed up the graph computation is to compute the graph on a downscaled image followed by an upscaling of the graph features. Single threaded computation of the graph on the  recycled concrete images after downscaling a by factor $4$ took only $15$ minutes with a peak memory usage of less than $1$ GiB. Since the computation on the downscaled images introduces approximation errors, this method is not recommended for computing the graphs of the real images. However, real image graphs usually need only be computed once and can then be used in the topological optimization for many synthetic images. In the case of synthetic images, we can utilize the fast computation of the graphs on the downscaled images. Since we optimize the graph topology afterwards, we can correct the approximation errors that worsen the topology. For example, when computing contacts between particles on the downscaled image, the upscaling of the edge features increases the edge weight of the contact from $1$ to $4$. During the optimization, this edge weight is likely to be decreased to $1$ again, correcting the error introduced through the graph computation.

\begin{figure*}[t]
    \centering
    \parbox{0.70\textwidth}{
    \subfloat[Particle graph]{
        \parbox{0.32\textwidth}{
            \centering
            \includegraphics[width=0.30\textwidth]{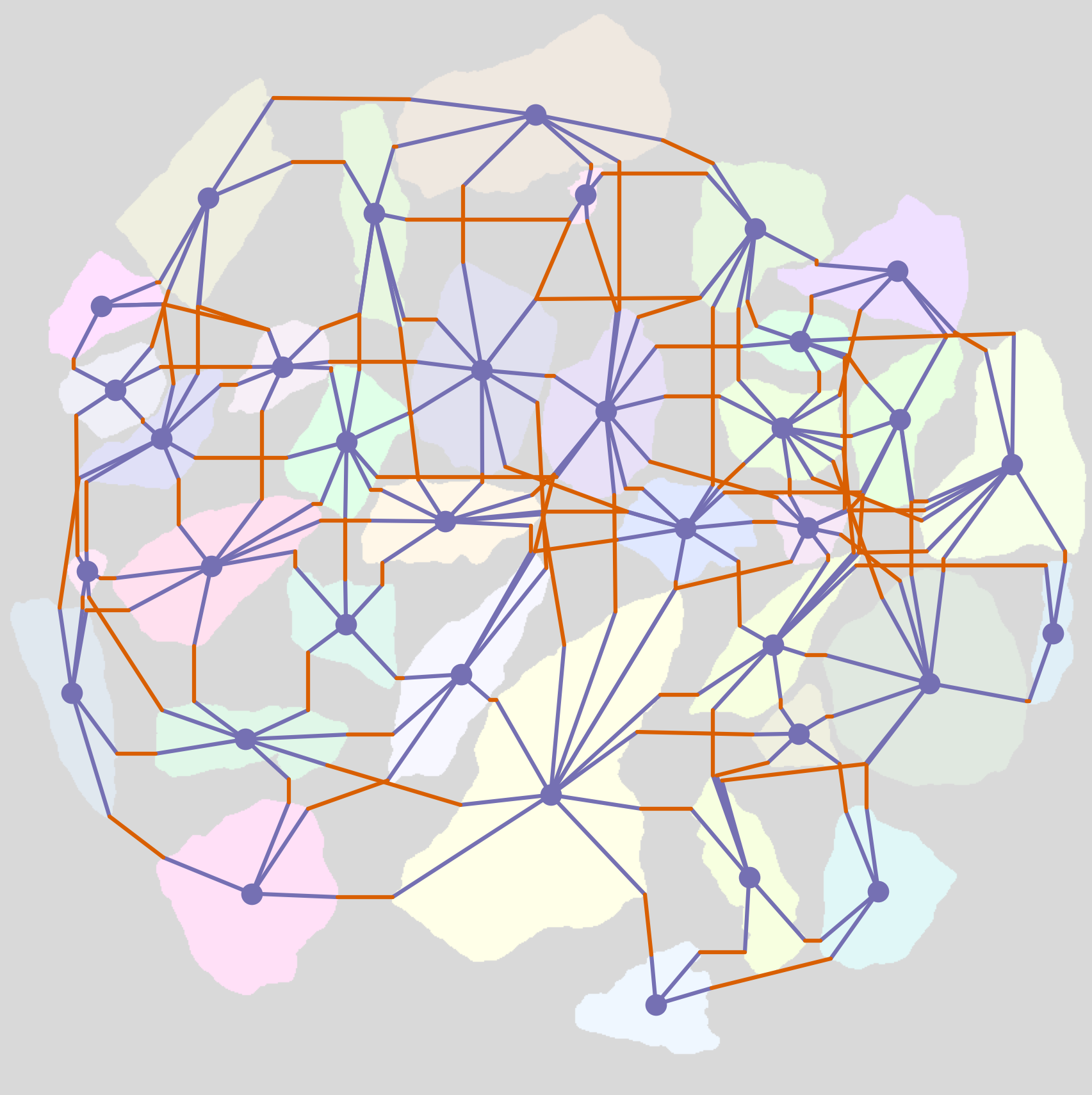}
        }
    }
    \hfill
    \subfloat[Reconstruction]{
        \parbox{0.32\textwidth}{
            \centering
            \includegraphics[width=0.30\textwidth]{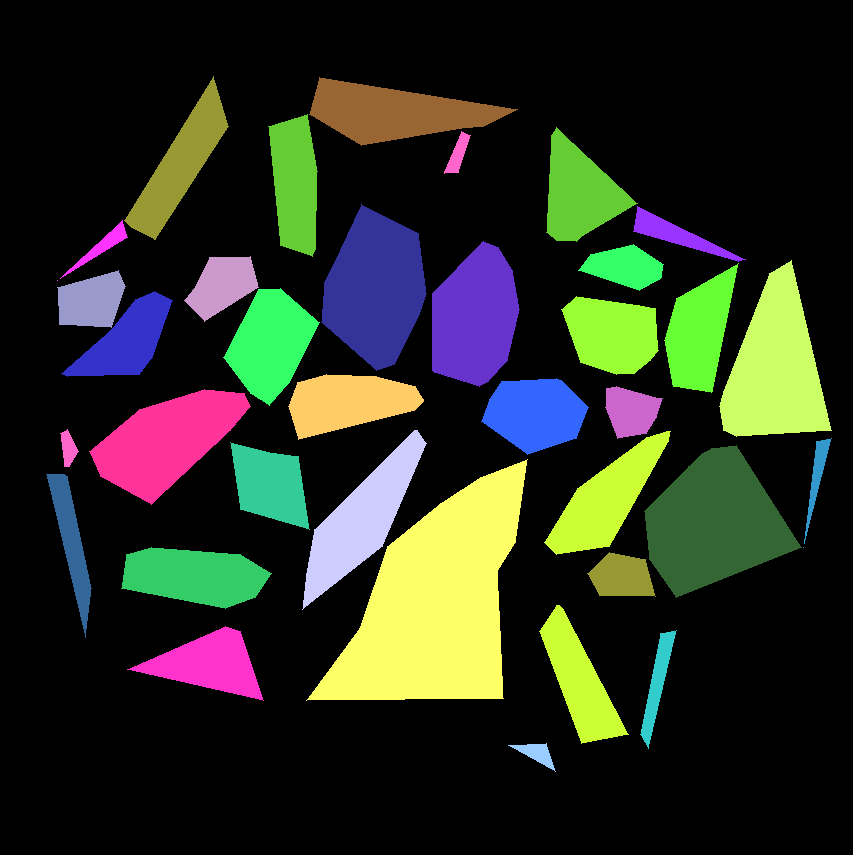}
        }
    }
    }

    \caption{A visualization of (a) the thresholded particle graph ($\tau = 150$) computed for a two-dimensional slice of a recycled concrete particle system and (b) the image reconstructed from the graph representation. For each edge in the displayed graph, the positions of the edge features correspond to the two transition points between the blue and orange segments. Poor shape reconstruction of the particles at the sample boundary is explained by missing information outside the cylindrical sample.}
    \label{Figure: Visualization Graph Representation}
\end{figure*}

In Figure~\ref{Figure: Visualization Graph Representation}, we visualize the idea of the graph representation in 2D on a slice taken from a computed tomography scan of recycled concrete particles.
The role of the node and edge features is clearly visible. In particular, connecting the edge features associated with a single particle and filling the interior of the resulting polygon approximately recovers the particle's shape and size, except for particles at the sample boundary. In the following, we extend this reconstruction to three-dimensional images.

\subsection{Image Reconstruction}
\label{Subsection: Image Reconstruction}

Given a three-dimensional image $\mathcal{I}$ and its (thresholded) particle graph $G$, we now define the reconstructed image $\widetilde{\mathcal{I}}$. We aggregate all points associated with a vertex $v \in V$ in a set $X_v$. More precisely, $X_v$ contains the center of mass $f_V(v)$ as well as, for all edges containing $v$, the point in $f_E(e)$ which belongs to $v$. The reconstructed particle $P_v$ is then defined as the convex hull of $X_v$, i.e. $P_v = \conv(X_v)$. If two reconstructed particles overlap, the voxels in their intersection are assigned to the particle with the larger label. The reconstructed image is then given by
\begin{equation}
    \label{Equation: Reconstructed Image Definition}
    \widetilde{\mathcal{I}}(p) = \max \left(\{0\} \cup \{v \in V \ \vert \ p \in P_v \} \right).
\end{equation}
The trimesh library~\cite{Trimesh2019} is used to discretize the convex hull of the particles into the voxel grid. The single threaded reconstruction of our recycled concrete data has a runtime of approximately $25$ minutes with a peak memory usage of $5$ to $7$ GiB depending on the size of the largest particle in the image. Since the reconstruction of the individual particles can be performed independently, a substantial speedup is expected from parallelization if shorter runtimes are required. The memory usage is not expected to increase linearly with the number of threads, since the reconstructed image (here approximately $4$~GiB) does not need to be copied for each worker.

The reconstruction can be followed by several postprocessing steps. Usually, we apply a morphological closing~\cite{Ohser_Schladitz2009ImageProcessingBook} with a ball of radius $1$ to close one voxel wide gaps between particles that result from the discretization of the reconstructed convex hull onto the voxel grid.

Moreover, the reconstruction via convex hulls yields very regular and smooth particle boundaries, which differ noticeably from the boundaries in the real images. To recover more realistic boundary irregularities, we can add simplex noise~\cite{Gustavson2005Simplex_Noise,Perlin2001Simplex_noise} to the particle boundary as a final postprocessing step. 

Simplex noise generates a spatially continuous noise function, typically in the interval $[-1,1]$. By scaling the coordinates of the boundary with a frequency parameter $f > 0$ before evaluating the noise function and multiplying the resulting noise by a magnitude parameter $M>0$, both the frequency and amplitude of the boundary perturbations can be controlled. More precisely, given a boundary point $x \in \mathbb{R}^3$ and a simplex noise function $n \colon \mathbb{R}^3 \rightarrow [-1,1]$, we perturb the boundary according to 
\begin{equation}
    x \mapsto x + M n(fx) \cdot \nu(x),    
\end{equation}
where $\nu(x)$ denotes the outward unit normal vector at x. Smaller values of $f$ and $M$ produce smoother boundaries, whereas larger values result in more irregular boundaries. The ability to control both the frequency and amplitude of the perturbations makes simplex noise well suited to our application.

\begin{figure*}[t]
    \centering
    \textbf{Comparison of Real and Reconstructed Images}

    \parbox{0.75\textwidth}{
    \subfloat[Real Image]{
            \parbox{0.23\textwidth}{
            \centering
            \includegraphics[width=0.18\textwidth]{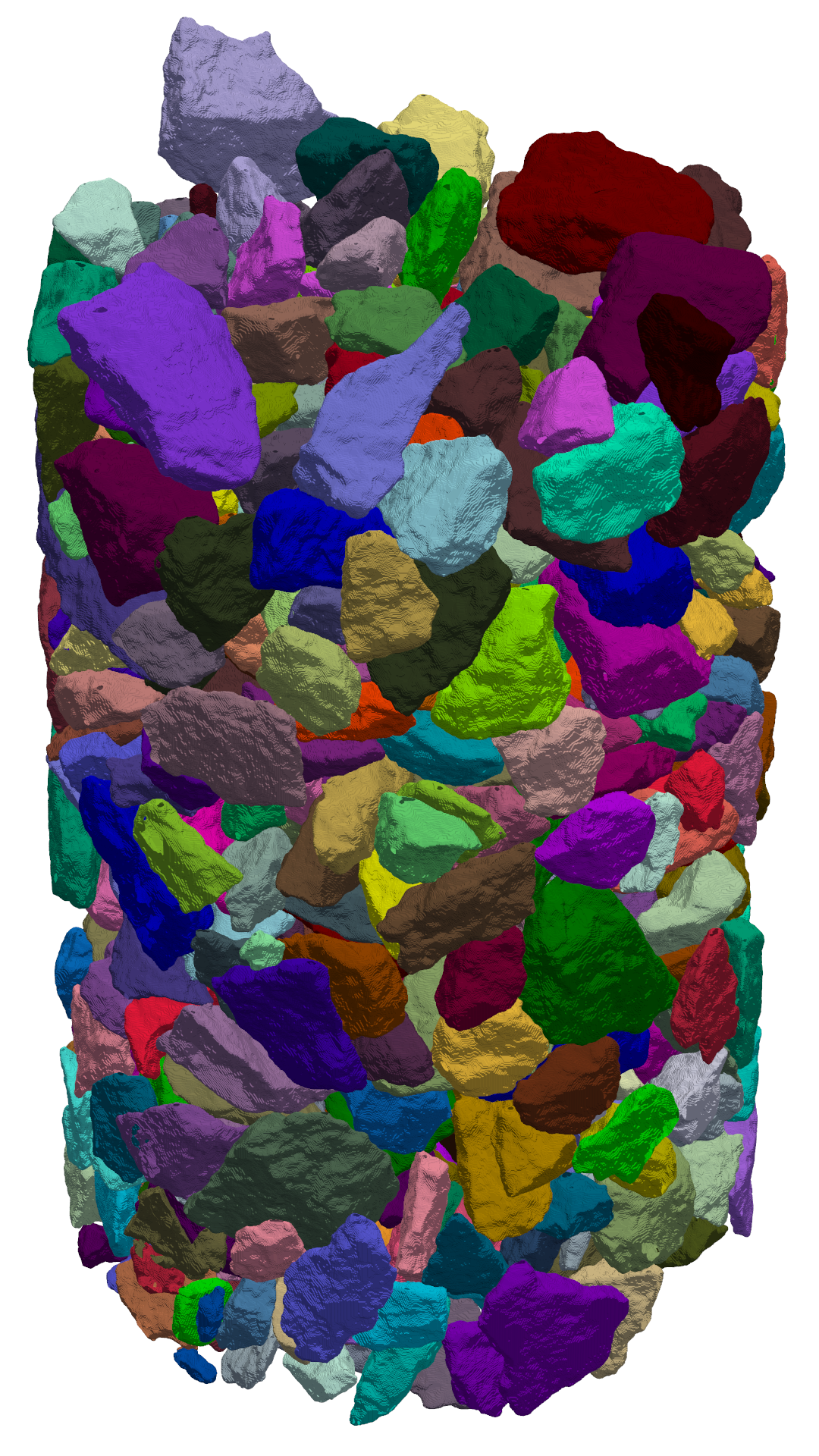}
            }
    }
    \hfill    
    \subfloat[Reconstruction]{
            \parbox{0.23\textwidth}{
            \centering
            \includegraphics[width=0.18\textwidth]{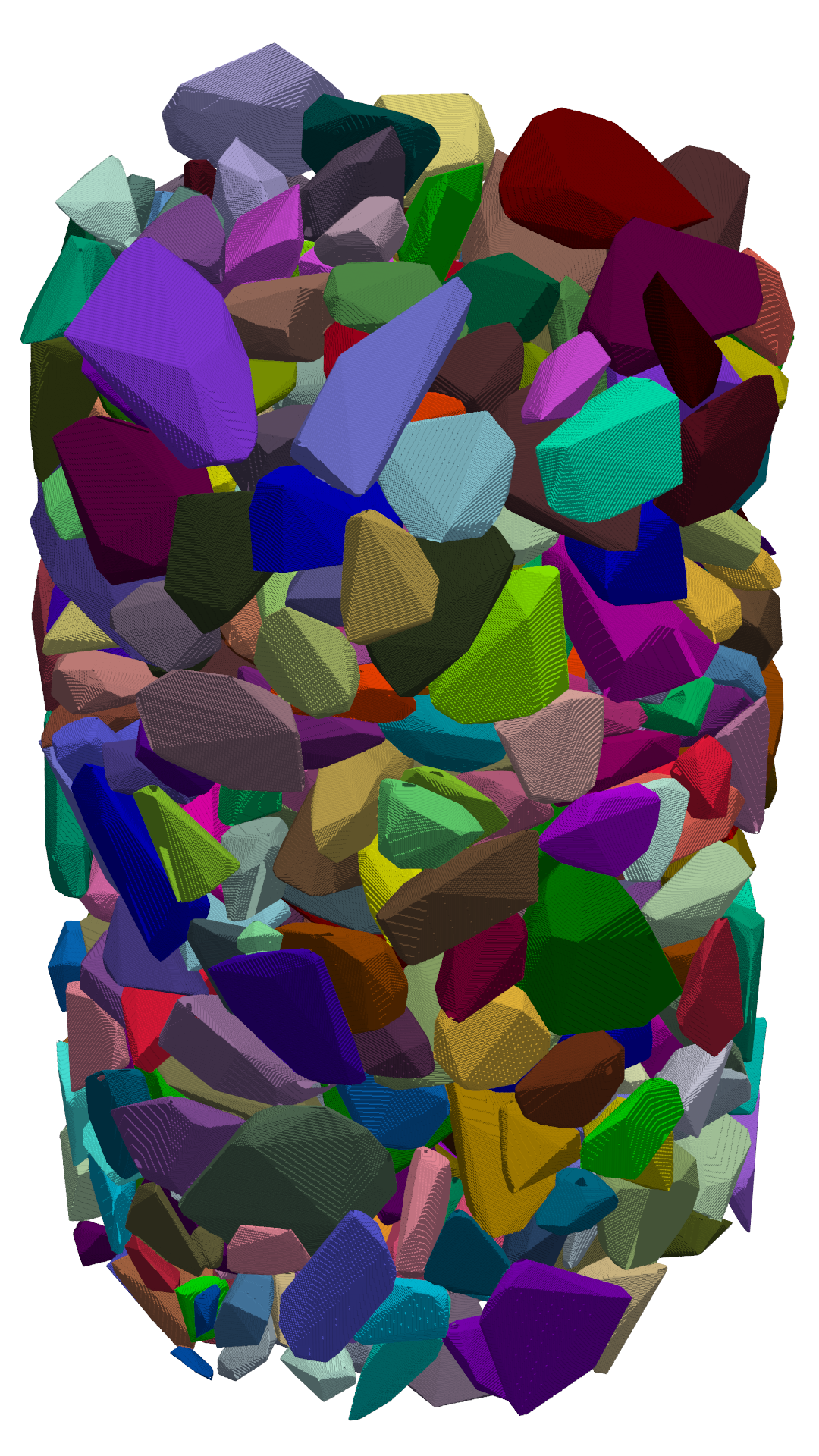}
            }
    }
    \hfill
    \subfloat[Single particle comparison]{
            \parbox{0.23\textwidth}{
            \centering
            \includegraphics[width=0.18\textwidth]{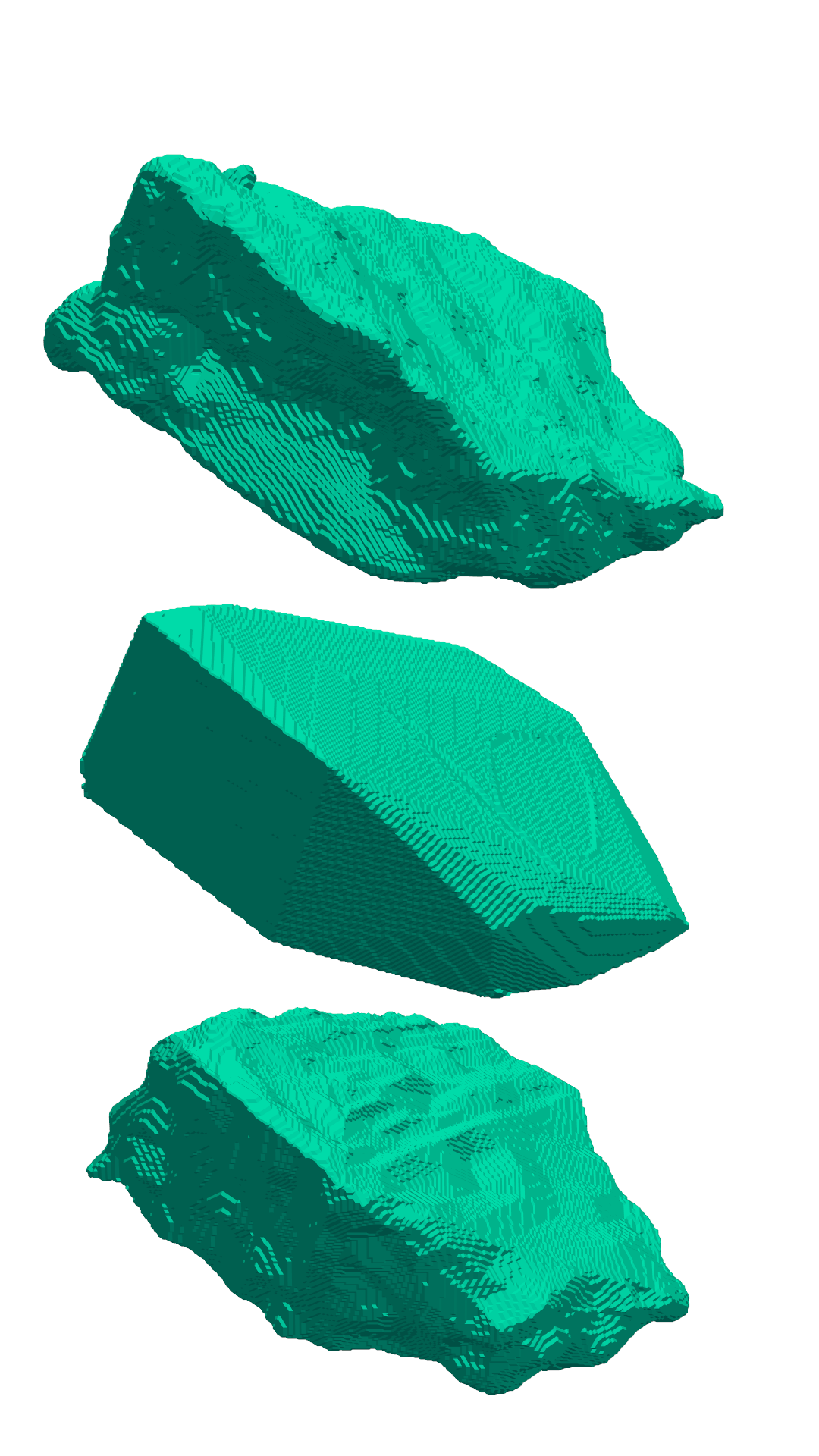}
            }
    }
    }
    \caption{Side-by-side comparison of (a) the real and (b) a reconstructed image of sample RC$1$. The reconstruction is computed from the thresholded particle graph with $\tau=400$. (c) shows a single particle comparison with the real particle, a reconstructed particle, and a reconstructed particle with simplex noise displayed from top to bottom. The simplex noise has frequency $f = 0.06$ and magnitude $M = 7$.}
    \label{Figure: Side-to-side comparison real and reconstructed image}
\end{figure*}

Figure~\ref{Figure: Side-to-side comparison real and reconstructed image} shows a side-by-side comparison of the real and reconstructed image, as well as a comparison of an individual particle and its corresponding reconstructed particle with and without simplex noise. Both particle shape and spatial arrangement are accurately recovered in the reconstruction. Compared with the previous 2D example, boundary artifacts are considerably less pronounced in the 3D reconstruction. Furthermore, the addition of simplex noise improves the visual appearance of the particle boundary. To quantitatively assess the reconstruction, we require a measure for evaluating its topological similarity, which is introduced in the next section.

\subsection{Relation between Image and Graph Topology}
\label{Subsection: Relation between Image and Graph Topology}

When carrying out the optimization in the graph domain instead of the image domain, we assume that the topological features in the graph domain can be related to the topological features in the image domain. For computing the persistence diagram on the particle graph, we define a weight function on the edges by 
\begin{equation}
    \label{Equation: Edge weight function of particle graph}
    w \colon E \rightarrow \mathbb{R}_+, e \mapsto \Vert u - v \Vert_1,    
\end{equation}
where $(u,v)=f_E(e)$. The persistence diagram of a graph is then computed using the VR-complex as introduced in Section~\ref{Subsections: Simplicial Homology on Graphs}.

When comparing the persistence diagrams computed on the image and on the graph, one has to keep in mind that we introduce slight differences and operate on different scales. Consider two particles with a distance of $d$ between their closest boundary voxels. In the graph, the two particles become connected at filtration value $d$, whereas in the image, they become connected at filtration value $\lceil \frac{d-1}{2}\rceil$. The $-1$ comes from the fact that the connectivity in the image is computed based on the boundaries of the voxels embedded in $\mathbb{R}^3$, while the distance for the graph is computed between their positions in the pixel grid $\Omega \subset \mathbb{Z}^3$. The division by $2$ comes from the fact that particles touch at half their distance when dilating the image. Since this scale difference is consistent throughout the construction, the graph persistence diagram can provide a consistent representation of the image persistence diagram despite the difference in scale.

The graph and its persistence diagram faithfully represent the connectivity structure of the original image. To see this, we note that thresholding the graph at $\tau = 1$ returns exactly the contact graph of the particle system that was discussed in Section~\ref{Subsection: Graph Construction}. If two particles are in contact with each other, a minimal unobstructed line segment is found exactly at the contact boundary voxels of the image and the weight of the corresponding edge is $1$. Hence, the connectivity in the graph and in the image coincide up to the difference in scale.

For higher dimensional features, we use two toy examples to demonstrate that the graph picks up topological features of the image. In the first toy example we place four cubes with a side length of $10$ into the corners of a $100 \times 100$ grid. This ($2$-dimensional) image then has four connected components which all merge at filtration time $40$. In the corresponding graph they merge at time $81$, which matches the scale difference discussed above. When the connected components merge in the image, they form a $1$-dimensional hole that closes at filtration time $80$. The graph correctly represents this hole as a topological feature in the persistence diagram at $(81,162)$. The scale difference, which was exact for $0$-dimensional features is also approximately true for this $1$-dimensional feature. 

The second toy example is constructed by placing a cube of side length $10$ at the center of each face of the cube defined by a $100 \times 100 \times 100$ grid. When computing the filtration of this image, the six cubes merge into one connected component at filtration time $36$ and form an octahedron like shape, resulting in a $2$-dimensional hole that disappears again at filtration time $40$. The graph also represents this $2$-dimensional hole with a point $(72,81)$ in the persistence diagram. Both in the image and graph topology no $1$-dimensional holes are detected when forming the triangle sides of the octahedron. In the graph, this is due to the Vietoris-Rips filtration, which cannot represent $1$-dimensional holes involving only $3$ nodes. In the image, this is due to the use of the $1$-norm in the distance transform. When using the $2$-norm instead, features with a short lifetime would be detected here. 

While the toy examples might give the impression that features are always detected, this is not true. The image persistence diagrams contain more points than the graph persistence diagrams, especially for the $1$-dimensional topology. This happens for multiple reasons. In the image, the same set of particles can generate multiple features with different parts of their boundary. In the graph, every particle pair is only considered once. Therefore, at most one of these features can be represented. Another possible reason is that two particles can have multiple contacts separated by concave areas. Through these contacts they can form $1$- and $2$-dimensional topological features. The graph representation cannot represent these features, since it needs at least $4$ nodes for a $1$-dimensional feature. In practice, this can happen a lot if the observed particles are not convex and have irregular boundaries. However, these features contributed by only $2$ particles usually have a short lifetime and can be considered topological noise.

\section{Topological Optimization}
\label{Section: Topological Optimization}
\subsection{Topological Loss Term and Differentiability}
\label{Subsection: Topological Loss Term and Differentiability}

The idea of our generative framework is as follows. First, we generate an initial synthetic particle system, then we match the topology of the corresponding particle graph with the topology of the particle graph of the real data. Finally, we reconstruct an image from the optimized synthetic graph. In this section, we focus on the topological optimization by defining a suitable loss term that can be minimized using gradient-based optimization.

Assume that we are given a thresholded particle graph $G$, computed from a synthetic image of a particle system with an arbitrary initial generation method. We compute the persistence diagrams of $G$ with the weight function defined in Eq.~\eqref{Equation: Edge weight function of particle graph} and denote them by $D_G^{(s)}$, where $s$ is the dimensionality of the topological features. The optimization objective is to match the topology of $G$ with the persistence diagrams of a particle graph computed from a real image of a particle system. 

Given a set of dimensions $S \subseteq \{0,1,2\}$, we want to optimize over persistence diagrams $\{D_G^{(s)}\}_{s \in S}$ and target diagrams $\{D^{(s)}\}_{s \in S}$. The logical candidate for the loss term of the optimization problem is a combination of Wasserstein distances for the different dimensions
\begin{equation}
    \label{Equation: Loss term}
    \mathcal{L}_{\operatorname{loss}}^{S} = \sum_{s \in S} \wasserstein_2^2(D_G^{(s)}, D^{(s)}).
\end{equation}
Using $\wasserstein_2^2$ instead of $\wasserstein_2$ is preferable as it avoids dealing with the square root in the gradients. This loss is not yet differentiable, since the $2$-Wasserstein contains an infimum in its definition. However, the $2$-Wasserstein distance can be rewritten in a differentiable form as
\begin{equation}
    \wasserstein_2^2(D_G^{(s)}, D^{(s)}) = \sum_{x \in D_G^{(s)}} \Vert x - T^{*}(x) \Vert_2^2,
\end{equation}
where $T^*$ is an optimal matching at which the infimum is attained. It is clear that such an optimal matching always exists, since there are only finitely many different matchings between the finite off-diagonal parts of the diagrams.

As discussed by Carriere et al.~\cite{carriere_al2026surveytopooptim}, the optimal matching is in general unique and therefore the $2$-Wasserstein distance is differentiable almost everywhere. However, the matching is an unstable quantity in the sense that slight perturbations in either $D_G^{(s)}$ or $D^{(s)}$ may lead to a significantly different optimal matching $T^*$ which consequently results in large changes in the corresponding gradients. This may cause instability in a gradient-based optimization framework. Instead of using the $2$-Wasserstein distance directly in the optimization, they propose to add an entropic regularization term based on the optimal transport problem, which improves the stability of the gradients. For a given regularization parameter $\gamma > 0$, we denote the entropic regularized $2$-Wasserstein distance by $\wasserstein_{2, \gamma}$. Details on the optimal transport theory and the entropic regularization term are provided in Appendix~\ref{Appendix: Entropic Regularization}.

Since $\wasserstein_{2, \gamma}$ provides stable gradients, it is a well suited option for the loss term in a gradient descent scheme. In particular, for a regularization parameter $\gamma > 0$, we replace the Wasserstein distances in Eq.~\eqref{Equation: Loss term} by their regularized approximations. The regularized loss term is given by
\begin{equation}
    \label{Equation: Regularized Loss term}
    \mathcal{L}_{\operatorname{loss}}^{S, \gamma} = \sum_{s \in S} \wasserstein_{2, \gamma}^2(D_G^{(s)}, D^{(s)}).
\end{equation}

To ensure end-to-end differentiability, the persistence diagrams $D_G^{(s)}$ must be differentiable w.r.t.\ the edge features. Since they are computed from a Vietoris-Rips filtration, each birth and death value corresponds to a value of the distance function from Eq.~\eqref{Equation: Distance Matrix of Vietoris-Rips}. Since the values of the distance function are either independent of the edge features or depend on the function $w$ from Eq.~\eqref{Equation: Edge weight function of particle graph}, it suffices to verify that $w$ is differentiable. While the $1$-norm is not differentiable when one of its components is $0$, a subgradient can be obtained by setting the corresponding component to $0$. Since subgradients can be employed in gradient-based optimization methods, this ensures applicability of our loss function. 

\subsection{Advantages of the Graph Representation}
\label{Subsection: Advantages of the Graph Representation}

Computing a graph and reconstructing an image introduces a domain gap that can both reduce or improve the quality of the synthetic images. Therefore, it is not obvious why the proposed graph representation should be preferred over operating directly on the image representation. In the following, we discuss the drawbacks of the image representation and the advantages of the proposed graph-based formulation within the optimization framework. 

To apply persistent homology to a labeled image and compute its persistence diagrams, a binarization is required. With this step, particle specific information is represented implicitly rather than explicitly. Moreover, even a binary image is not well suited for a gradient descent algorithm, as it is unclear how to define meaningful gradient updates. The standard approach to mitigate this issue is by relaxing the image to continuous values between $0$ and $1$, which in turn requires thresholding to recover a binary representation for the computation of the persistence diagrams.

One of the key advantages of synthetic data over real data is, that a perfect segmentation is known. However, the implicit representation of particle specific information removes this advantage since the data is no longer explicitly labeled and segmentation must be performed after the optimization. Moreover, the implicit representation allows particles to merge or the creation of new particles during the optimization. While this may still yield the desired topological properties, it can drastically alter other characteristics, such as the particle size distribution.

Another drawback of operating on the image domain is the high computational cost. Three-dimensional $\mu$CT images typically contain a large number of voxels (e.g., the images of recycled concrete considered in this work have a size of approximately $800 \times 800 \times 1600$ voxels). Computing a persistence diagram on such an image is computationally very expensive. For the recycled concrete data we measured computation times of approximately $25$ minutes with a peak memory usage of approximately $135$ GiB using the CubicalRipser~\cite{CubicalRipser2020} library. Downsampling the image substantially reduces the computational cost. For example, downsampling the image by a factor of $4$ in all dimensions reduces the computation time to approximately $18$ seconds with a peak memory usage of $3$ GiB. While this is much faster, it is still computationally expensive for use in a gradient descent algorithm, which requires many iterations. Furthermore, particle systems with smaller particles might lose too much information when using such an aggressive downsampling factor.

The graph representation mitigates all of these issues. Since the edge features are continuous, gradient-based optimization can be applied directly with huge computational benefits. For the recycled concrete data with $500$ to $600$ particles computing a persistence diagram took less than $0.4$ seconds with a peak memory usage of $0.8$ GiB, using the single-threaded version of the giotto-ph~\cite{burella2021giottoph} library. Furthermore, the graph preserves the explicit particle information, and the reconstructed image is inherently segmented. The explicit particle information also allows the gradient updates to be interpreted as a local deformation of the particle boundary in the neighborhood of the moved edge features. In particular, particles cannot merge, and no new particles can be created during the optimization algorithm.

A naturally arising question is why we propose a graph-based representation rather than using a marked point cloud. The reconstructed image does indeed only depend on the points stored in the edge features and is independent of the actual edge set $E$ (cf.\ Eq.~\eqref{Equation: Reconstructed Image Definition}). By setting the distance of two points with the same mark to $0$ and applying the VR filtration to the point cloud, this would provide a reasonable representation of the topology of the particle system. 

However, this approach is computationally infeasible in practice. The cost of computing persistence diagrams via the Vietoris-Rips filtration depends on the number of simplices in the filtration, which grows rapidly with the number of input points. In our setting, the particle graphs contain between $21{,}000$ and $25{,}000$ edges, corresponding to a point cloud of $42{,}000$ to $50{,}000$ points. Even when using a threshold on the filtration values of the simplices, the computational cost is prohibitively large. 

Rather than representing each point separately, our proposed graph representation aggregates all points belonging to the same particle into a single graph node and encodes their geometry in the weight function $w$. As a result, a configuration of over $40{,}000$ points is encoded as a graph with $500$ to $600$ nodes. This substantially reduces the size of the induced VR complexes in the filtration and, in turn, the computational cost of persistent homology.

\section{Experimental Validation}
\label{Section: Experimental Validation}
\subsection{\texorpdfstring{$\mu$CT}{CT} Images of Particle Systems}
\label{Subsection: muCT Images of Particle Systems}

\begin{figure*}[t]
    \centering
    $\boldsymbol{\mu}$\textbf{CT Images of Recycled Concrete}

    \parbox{0.8\textwidth}{
    \subfloat[RC$1$]{
        \includegraphics[width=0.18\textwidth]{S-S2_4-1.png}
    }
    \hfill
    \subfloat[RC$2$]{
        \includegraphics[width=0.18\textwidth]{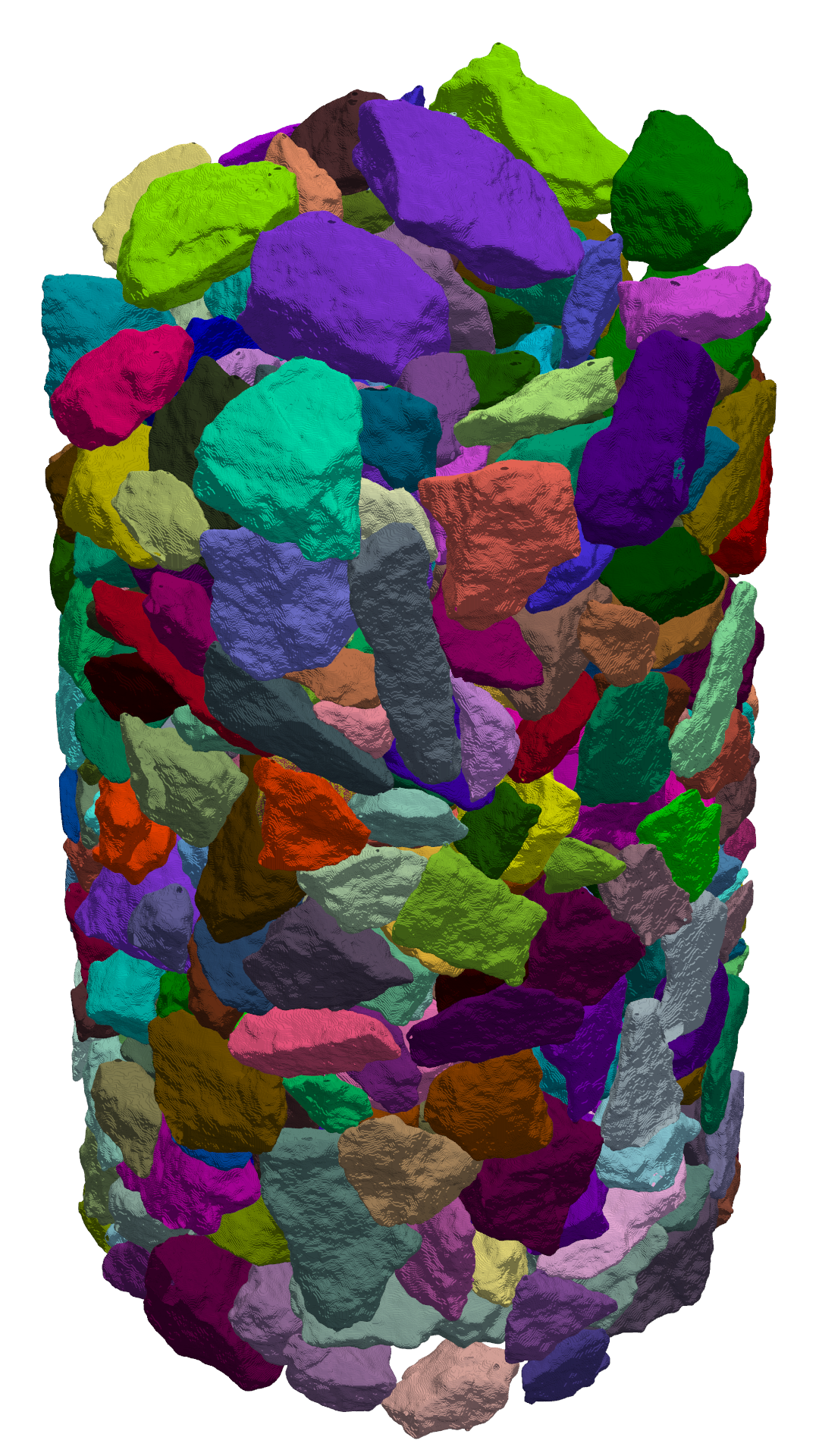}
    }
    \hfill
    \subfloat[RC$3$]{
        \includegraphics[width=0.18\textwidth]{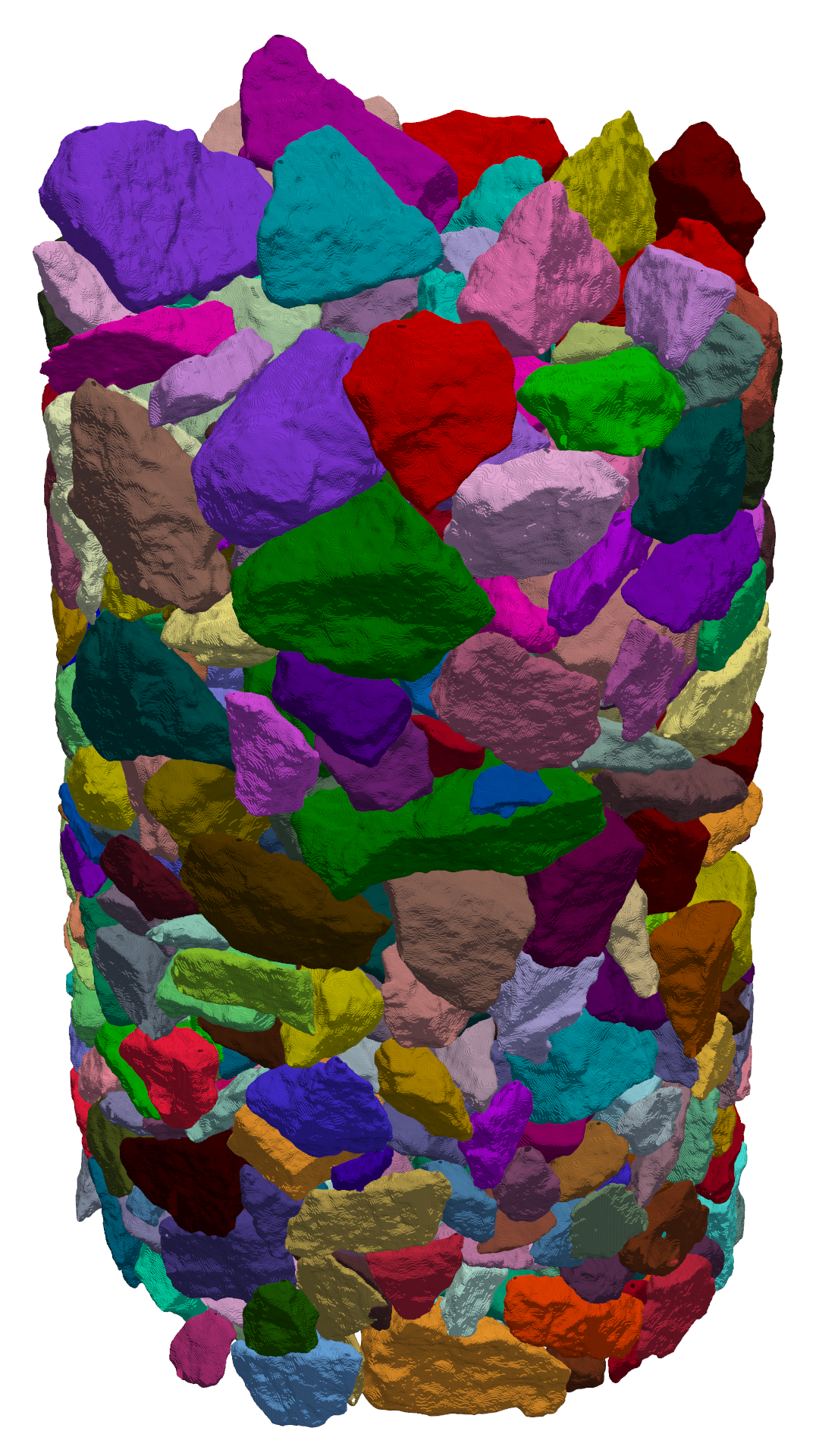}
    }
    \hfill
    \subfloat[RC$4$]{
        \includegraphics[width=0.18\textwidth]{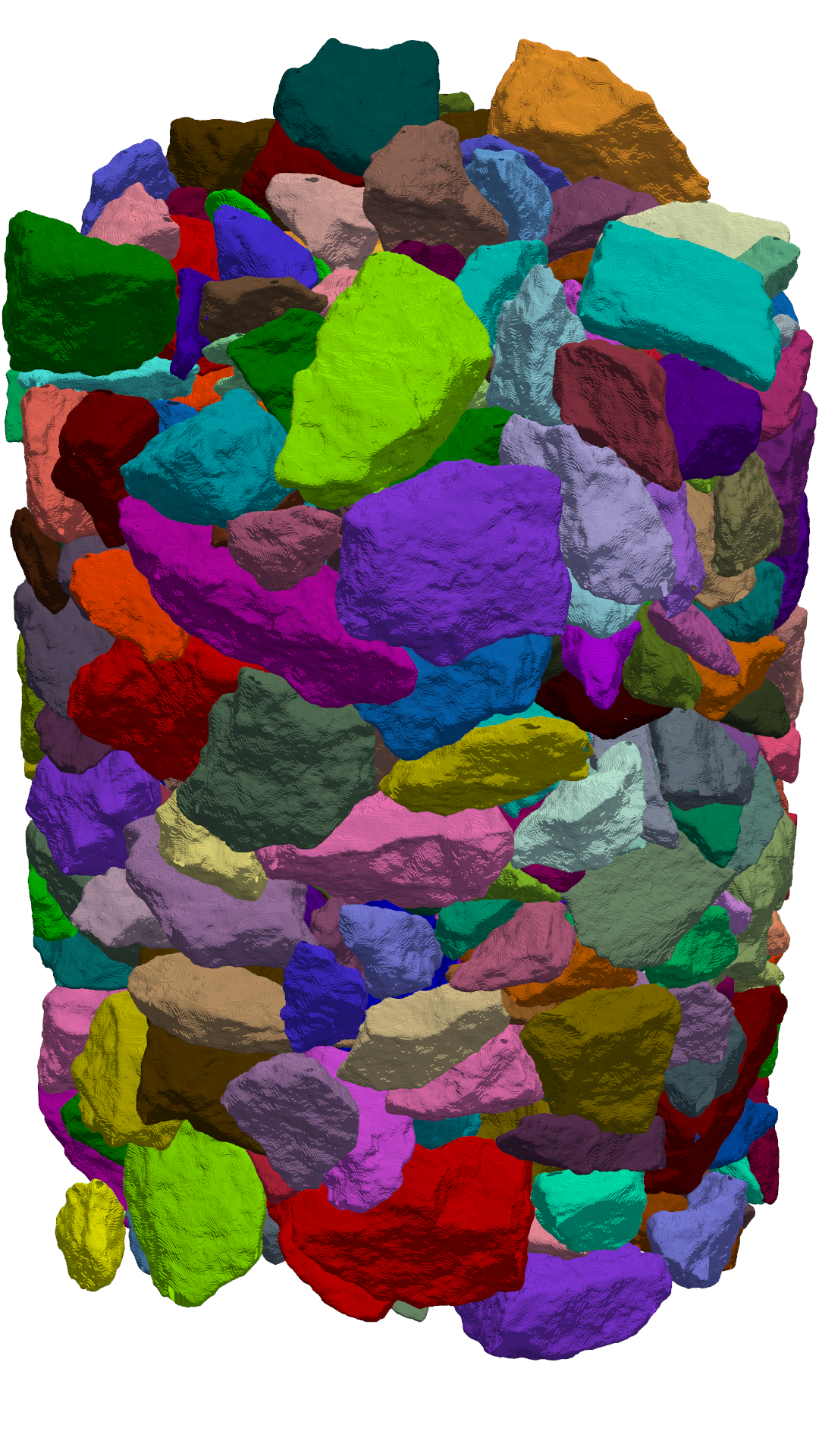}
    }
    }
    \caption{Visualization of four labeled $\mu$CT images of recycled concrete scanned as packed beds in a cylindrical scanning container.}
    \label{Figure: Recycled Concrete Images}
\end{figure*}

In this study we consider a set of four three-dimensional X-ray micro-computed tomography ($\mu$CT) images of recycled concrete (RC) particles. The particle systems were scanned as packed beds in a cylindrical scanning container. The X-ray projections are reconstructed using VGSTUDIO 3.2 (Volume Graphics) by the Feldkamp--David--Kress algorithm~\cite{Feldkamp_al1984ReconstructionCTImage}. After several postprocessing steps, the image is converted into a binary image using Otsu's method~\cite{Otsu1979Thresholding}. The particles are then segmented using the watershed transform~\cite{Vincent_and_Soille1991Watershed}. To remove the separation boundaries introduced by the watershed transform, we subsequently apply a morphological closing with a ball of radius $1$ voxel. The resulting particle systems are visualized in Figure~\ref{Figure: Recycled Concrete Images}. Further details on the image data and the processing steps can be found in~\cite{Burgmann_al2022DetailsOfData}.
\subsection{Validation of the Reconstruction Algorithm}
\label{Subsection: Validation of the Reconstruction Algorithm}

For method validation, we check that topological and geometric characteristics of the reconstructed particles reproduce the characteristics observed in the four images of recycled concrete. For every image we compute the thresholded particle graph with threshold $\tau=400$. From that graph we compute the reconstructed image following Eq.~\eqref{Equation: Reconstructed Image Definition} and postprocess it with a morphological closing of radius $1$. 

We report five metrics in total, two pixelwise and three topological metrics. For measuring the pixelwise accuracy, we report the Dice score~\cite{Mohapatra_al2023dice_in_survey} which, for two sets $X$ and $Y$ is given by
\begin{equation}
    D(X,Y) = \frac{2 \vert X \cap Y \vert}{\vert X \vert + \vert Y \vert}.
\end{equation}
We report the mean of the Dice scores of the individual particles in the labeled image, and the Dice score of the full image which takes the entire foreground as the ground-truth set. For the topological metrics we compute the persistence diagrams of the real and reconstructed images as described in Section~\ref{Subsection: Cubical Homology on Images}. For the $1$- and $2$-dimensional persistence diagrams we remove noisy features with persistence $2$ or less from the corresponding persistence diagrams, i.e. points $(b,d)$ satisfying $d-b \leq 2$. Then we compute the $2$-Wasserstein distances. As the Wasserstein distance is not an absolute measure, we instead report normalized $2$-Wasserstein distances for the $1$- and $2$-dimensional diagrams obtained by dividing each distance by the $2$-Wasserstein distance between the persistence diagram of the real image and the empty diagram (containing only the diagonal). This allows the interpretation of the value as the fraction of total topological change during the reconstruction. For the $0$-dimensional diagrams we neither do the filtering nor the normalization. Since all binarized real images in our dataset consist of a single connected component, their $0$-dimensional persistence diagrams are empty and the normalization is therefore not well defined. The resulting metrics are denoted by $H_s$ with $s \in \{0,1,2\}$ denoting the dimensionality of the persistence diagrams.

\begin{table}[t]
    \centering
    \textbf{Reconstruction Quality}
    \vspace{.2em}

    {\small
    \begin{tabular}{lccccc}
        \toprule
        Image & Mean Dice $\uparrow$ & Dice $\uparrow$ & $H_0 \downarrow$ & $H_1 \downarrow$ & $H_2 \downarrow$ \\
        \midrule
        RC$1$ & $0.867 \pm 0.067$ & $0.873$ & $0$ & $0.166$ & $0.273$ \\
        RC$2$ & $0.874 \pm 0.059$ & $0.879$ & $0$ & $0.188$ & $0.258$ \\
        RC$3$ & $0.874 \pm 0.066$ & $0.876$ & $0$ & $0.200$ & $0.268$ \\
        RC$4$ & $0.872 \pm 0.066$ & $0.875$ & $0$ & $0.186$ & $0.264$ \\
        \bottomrule
    \end{tabular}
}

\caption{Reconstruction results of four images of recycled concrete. We report two Dice score metrics and three Wasserstein distances ($\uparrow$: higher is better, $\downarrow$: lower is better).}
\label{Table: Experiment Reconstruction Metrics}

\end{table}

The results are reported in Table~\ref{Table: Experiment Reconstruction Metrics}. The observed Dice scores are around $0.87$, indicating a high degree of overlap between the particles in the reconstructed and the real image. Moreover, the mean Dice score suggests that this performance is not driven by a few large particles dominating the metric, but rather by a consistent overlap across individual reconstructed particles. We also observe a very strong performance across the topological metrics. The $0$-dimensional topology shows that the reconstructed image again contains a single connected component. In the $1$-dimensional topology approximately $82\%$ of the total topology is preserved through the reconstruction, while the corresponding value for the $2$-dimensional topology is approximately $74\%$. 

The reported values correspond to the reconstruction without the addition of simplex noise. We also tested the effect of incorporating simplex noise and observed that the Dice scores consistently decreased by approximately $0.01$. The topological metrics varied only slightly, being consistent in $H_0$, improving for $H_1$ and deteriorating for $H_2$. Overall, simplex noise does not have a consistent positive or negative effect on the reconstruction quality. Therefore, it is omitted from subsequent evaluations.

The observations in this section are robust w.r.t.\ the threshold value $\tau$. We evaluated the reconstructions for thresholds between $\tau=200$ and $\tau=600$. When increasing the threshold, the Dice scores improved from $0.84$ to $0.89$, while the topological metrics showed no clear trend but varied only slightly for both the $1$- and $2$-dimensional topology. Thus, across the tested range, the conclusions regarding reconstruction quality are mostly insensitive to the choice of $\tau$. The main effect of varying $\tau$ is observed in the volume fraction in the resulting reconstructed image. A higher threshold results in more points being used in the convex hull reconstruction, which leads to an increase in volume fraction. Therefore, $\tau$ should be chosen to reproduce the volume fraction of the original image as accurately as possible. For the recycled concrete images considered here, $\tau = 400$ provides the closest match.
\subsection{Synthetic Images of Concrete}
\label{Subsection: Synthetic Images of Concrete}

We now evaluate the reconstruction and topological optimization on synthetic images. For each of the four real concrete images, we generated $16$ synthetic images with the random sequential adsorption algorithm (RSA). Details of RSA are given in Appendix~\ref{Appendix: RSA}. Then, we optimized the topology of the corresponding thresholded particle graphs with $\tau=400$. The optimization used the regularized loss term from Eq.~\eqref{Equation: Regularized Loss term} for different $S \subseteq \{0,1,2\}$ and the Adam optimizer~\cite{Kingma_and_Ba2015AdamOptimization} for $500$ iterations. Implementation details and hyperparameters of the optimization can be found in Appendix~\ref{Appendix: Experiment Details}. Finally, we reconstructed images from the graphs. Before evaluation we applied a morphological closing of radius $1$ to both the RSA and the reconstructed images and refrained from adding simplex noise to the reconstruction. We compare the optimized reconstructed images to the real and the RSA images in terms of the number of particles, the volume fraction, the Wasserstein distances, the particle volume and surface area distribution, and the geometric tortuosity.

\begin{table}[t]
    \centering
    \textbf{Number of Particles}
    \vspace{.2em}

    {\small
    \begin{tabular}{lcccc}
        \toprule
        Image Type & RC$1$ & RC$2$ & RC$3$ & RC$4$ \\
        \midrule

        Real Image & $510$ & $517$ & $579$ & $504$ \\
        RSA & $510$ & $513.438$ & $575.125$ & $503.875$ \\
        \bottomrule
    \end{tabular}
    }
    
    \caption{Number of particles in the real images and average number of particles placed in the RSA image.}
    \label{Table: Number of particles in orig and RSA}
\end{table}

\begin{table}[t]
    \centering
    \textbf{Volume Fraction}
    \vspace{.2em}

    {\small
    \begin{tabular}{lcccc}
        \toprule
        Image Type & RC$1$ & RC$2$ & RC$3$ & RC$4$ \\
        \midrule
        Real Image & $0.410$ & $0.454$ & $0.472$ & $0.421$ \\
        RSA & $0.410$ & $0.455$ & $0.472$ & $0.422$ \\
        Reconstruction & $0.402$ & $0.446$ & $0.469$ & $0.415$ \\
        $S=\{0\}$ & $0.405$ & $0.448$ & $\mathbf{0.471}$ & $0.417$ \\
        $S=\{1\}$ & $0.415$ & $0.458$ & $0.482$ & $0.428$ \\
        $S=\{2\}$ & $0.406$ & $0.452$ & $0.476$ & $\mathbf{0.422}$ \\
        $S=\{0,1\}$ & $0.416$ & $0.458$ & $0.482$ & $0.429$ \\
        $S=\{0,2\}$ & $\mathbf{0.409}$ & $\mathbf{0.454}$ & $0.478$ & $0.425$ \\
        $S=\{1,2\}$ & $0.418$ & $0.462$ & $0.487$ & $0.434$ \\
        $S=\{0,1,2\}$ & $0.418$ & $0.462$ & $0.487$ & $0.434$ \\
        \bottomrule
    \end{tabular}
    }

    \caption{Volume fraction of particles inside the cylindrical scanning container. We report the volume fraction of the real images and the mean volume fractions of the $16$ synthetic images. The best fit is highlighted in bold.}
    \label{Table: Volume Fraction of Real and Synthetic Images}
\end{table}

The maximal optimization runtime across all $64$ synthetic images and combinations of $S$ is less than $5$ minutes. Furthermore, if the $2$-dimensional topology is not used in the optimization, the persistence diagram computation is significantly faster and the runtime decreases to less than $100$ seconds. If only the $0$-dimensional topology is used, the optimization is fastest, with the optimized graph being computed in less than $50$ seconds.

Since the particle systems in the real images are relatively dense, it is not guaranteed that the RSA algorithm is able to place all particles. Table~\ref{Table: Number of particles in orig and RSA} reports the number of particles in the real images together with the average number of particles placed in the corresponding RSA images. We observe that all particles could be placed in RC$1$, while in RC$4$ only $2$ particles across all $16$ synthetic images could not be placed. For RC$2$ and RC$3$, on average around $3$ to $4$ particles per image are missing. Nevertheless, the volume fractions of the real and RSA images are nearly identical, see Table~\ref{Table: Volume Fraction of Real and Synthetic Images}. In fact, the volume fraction of the RSA images is sometimes even slightly larger than that of the real images which is caused by  the morphological closing applied to the RSA images. Furthermore, we observe that the reconstruction using the unoptimized graph leads to a slight loss of volume fraction. Optimizing the graph increases the volume fraction of the corresponding reconstructed images, with most of this increase coming from the $1$-dimensional topology.

\begin{table*}[t]
    \centering
    $\mathbf{2}$\textbf{-Wasserstein Distances}
    \vspace{.2em}

    {\small
    \begin{tabular}{lccc ccc ccc ccc}
        \toprule
        
        & \multicolumn{3}{c}{RC$1$} & \multicolumn{3}{c}{RC$2$} & \multicolumn{3}{c}{RC$3$} & \multicolumn{3}{c}{RC$4$} \\
        \cmidrule(lr){2-4} \cmidrule(lr){5-7} \cmidrule(lr){8-10} \cmidrule(lr){11-13}
        
        Image Type
        & $H_0$ & $H_1$ & $H_2$
        & $H_0$ & $H_1$ & $H_2$
        & $H_0$ & $H_1$ & $H_2$
        & $H_0$ & $H_1$ & $H_2$ \\
        \midrule 

        RSA &
        $14.451$ & $0.505$ & $0.531$ &
        $\phantom{1}6.553$ & $0.430$ & $0.483$ &
        $\phantom{1}5.650$ & $0.450$ & $0.508$ &
        $\phantom{1}9.797$ & $0.551$ & $0.591$ \\

        Reconstruction &
        $12.077$ & $0.533$ & $0.501$ &
        $\phantom{1}5.621$ & $0.472$ & $0.454$ &
        $\phantom{1}5.135$ & $0.483$ & $0.480$ &
        $\phantom{1}7.540$ & $0.581$ & $0.560$ \\

        $S=\{0\}$ &
        $\phantom{1}0.663$ & $0.525$ & $0.500$ &
        $\phantom{1}\mathbf{0.044}$ & $0.464$ & $0.452$ &
        $\phantom{1}\mathbf{0.000}$ & $0.475$ & $0.476$ &
        $\phantom{1}\mathbf{0.000}$ & $0.573$ & $0.557$ \\

        $S=\{1\}$ &
        $\phantom{1}8.934$ & $0.421$ & $0.475$ &
        $\phantom{1}4.019$ & $0.369$ & $0.419$ & 
        $\phantom{1}3.751$ & $0.379$ & $0.443$ &
        $\phantom{1}5.166$ & $0.470$ & $0.517$ \\

        $S=\{2\}$ &
        $11.857$ & $0.523$ & $0.483$ &
        $\phantom{1}5.487$ & $0.449$ & $0.416$ &
        $\phantom{1}4.980$ & $0.458$ & $0.436$ &
        $\phantom{1}7.330$ & $0.563$ & $0.506$ \\

        $S=\{0,1\}$ &
        $\phantom{1}\mathbf{0.447}$ & $0.419$ & $0.474$ &
        $\phantom{1}\mathbf{0.044}$ & $0.370$ & $0.419$ &
        $\phantom{1}\mathbf{0.000}$ & $0.380$ & $0.443$ &
        $\phantom{1}0.044$ & $0.466$ & $0.517$ \\

        $S=\{0,2\}$ &
        $\phantom{1}0.751$ & $0.515$ & $0.481$ &
        $\phantom{1}0.088$ & $0.441$ & $0.413$ & 
        $\phantom{1}0.088$ & $0.448$ & $0.433$ &
        $\phantom{1}0.044$ & $0.555$ & $0.503$ \\

        $S=\{1,2\}$ &
        $\phantom{1}8.984$ & $0.417$ & $\mathbf{0.463}$ &
        $\phantom{1}3.894$ & $\mathbf{0.359}$ & $0.390$ &
        $\phantom{1}3.672$ & $\mathbf{0.369}$ & $0.411$ &
        $\phantom{1}5.064$ & $0.460$ & $0.470$ \\

        $S=\{0,1,2\}$ &
        $\phantom{1}1.063$ & $\mathbf{0.415}$ & $\mathbf{0.463}$ &
        $\phantom{1}0.088$ & $0.360$ & $\mathbf{0.389}$ &
        $\phantom{1}0.313$ & $0.370$ & $\mathbf{0.410}$ &
        $\phantom{1}0.044$ & $\mathbf{0.457}$ & $\mathbf{0.469}$ \\

        \bottomrule
    \end{tabular}
    }
    
    \caption{Mean $2$-Wasserstein distances between the synthetic images and the corresponding target image. Best results are highlighted in bold (lower is better).}
    \label{Table: 1-Wasserstein distance synthetic to real image (normalized)}

\end{table*}

The main objective of the optimization is to decrease the topological difference between the real and synthetic images. We report the same topological metrics $H_0, H_1, H_2$, computed from the $2$-Wasserstein distances, as defined in Section~\ref{Subsection: Validation of the Reconstruction Algorithm}. The results are shown in Table~\ref{Table: 1-Wasserstein distance synthetic to real image (normalized)}. We observe that the domain gap introduced by computing the reconstructed image without any optimization can have both a positive or negative effect on the topology. In particular, we observe a positive effect across all four images in $H_0$ and $H_2$ and a negative effect in $H_1$. However, the differences are relatively small such that the reconstructions of the optimized graphs outperform the RSA images consistently in all three dimensions.

A first observation is that optimizing w.r.t. a given dimension consistently improves the corresponding metric in the resulting image. In particular, the best observed values for each dimension are always achieved by an optimization that uses the same dimension. Furthermore, we observe that incorporating additional dimensions usually helps, e.g.\ the best observed values of $H_1$ and $H_2$ are always attained when both of them are in $S$. However, $H_0$ is usually worse if additional dimensions are included during the optimization. 

We now consider the best observed values for $H_0, H_1, H_2$ individually and compare them with the values of the synthetic RSA images and unoptimized reconstructions. For $H_0$ a perfect agreement is achieved for RC$3$ and RC$4$ across all $16$ synthetic images, while the errors are reduced by over $95\%$ for RC$1$ and RC$2$. For $H_1$, the optimized images improve by between $21\%$ and $24\%$ relative to the unoptimized reconstructions and by between $16\%$ and $18\%$ relative to the RSA images. For $H_2$, the improvement relative to the unoptimized reconstructions ranges from approximately $7\%$ for RC$1$ to between $14\%$ and $17\%$ for the other three images. Compared with the RSA images, the improvements increase to approximately $13\%$ for RC$1$ and about $20\%$ for the other three images. Overall, these results demonstrate that optimization on the graph representation consistently improves the image topology across all three dimensions.

\begin{figure*}[t]
    \centering

    \includegraphics[width=0.78\textwidth]{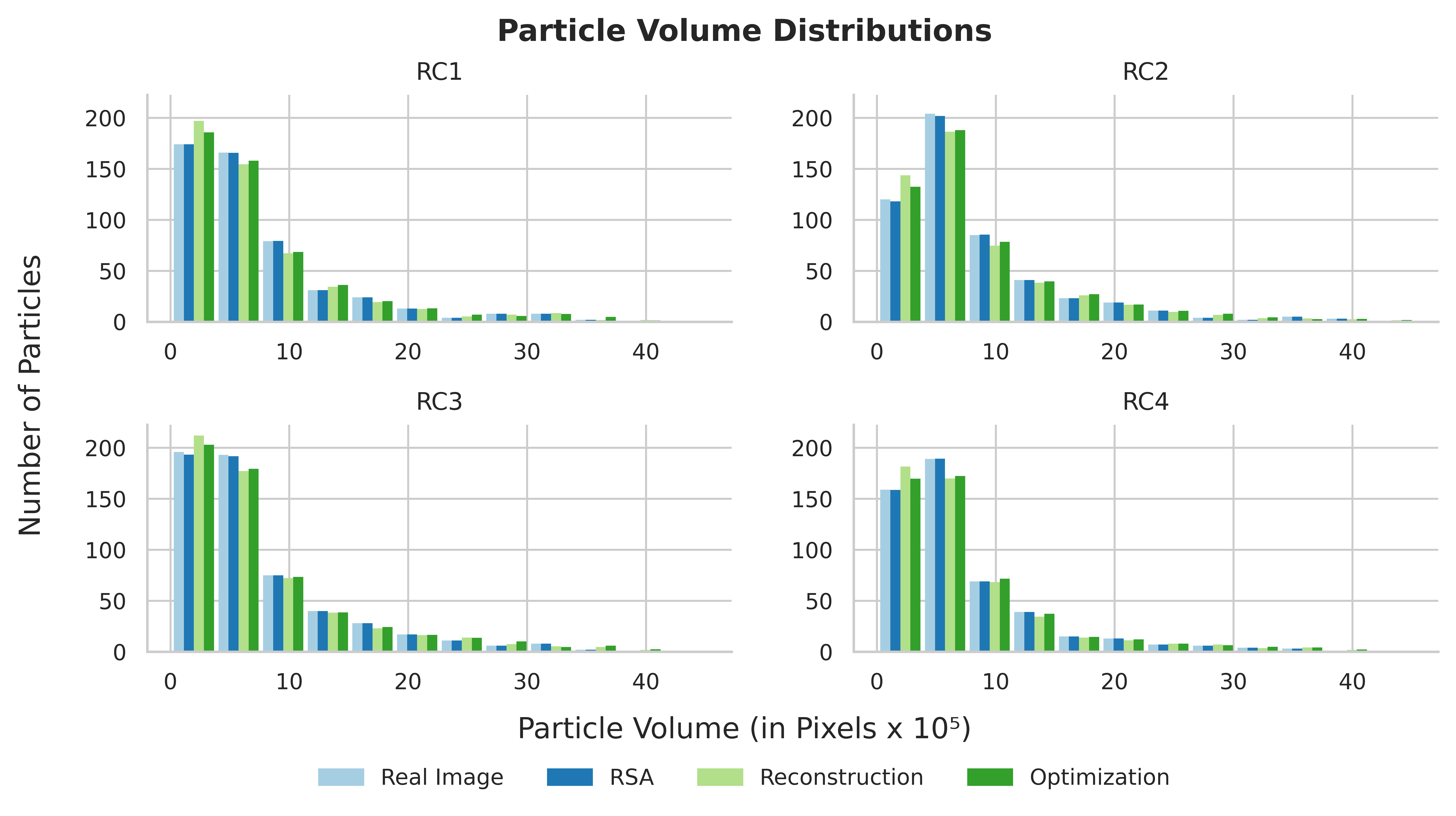}
    \caption{Comparison of the particle volume distributions between the real images, the synthetic RSA images, their reconstructions from the graph, and the reconstructed images of the graphs optimized on $S = \{0,1,2\}$.
    \label{Figure: Histograms Particle Volume Distribution}
    }

\end{figure*}

\begin{figure*}[t]
    \centering

    \includegraphics[width=0.78\textwidth]{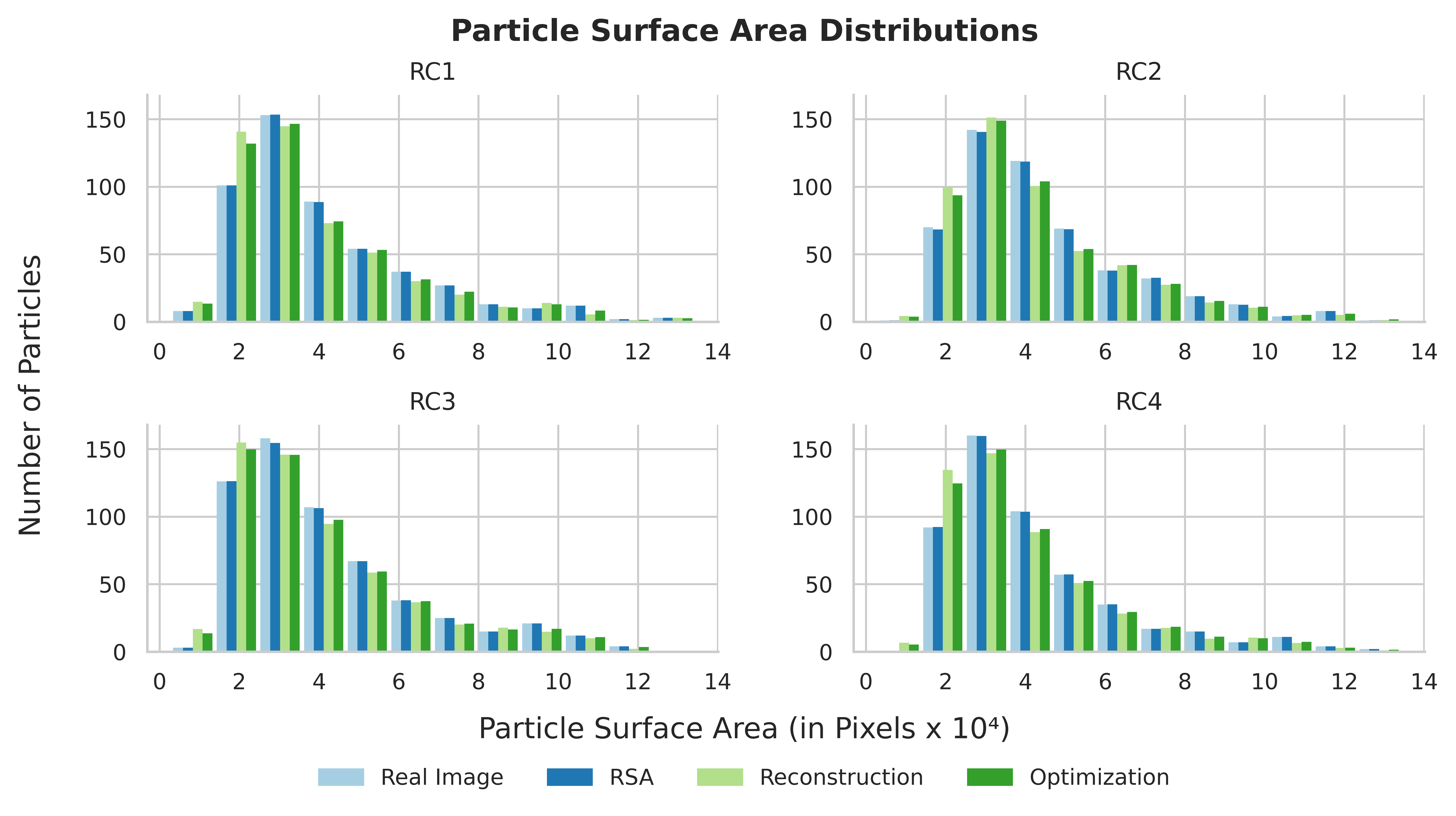}
    \caption{Comparison of the particle surface area distributions between the real images, the synthetic RSA images, their reconstructions from the graph, and the reconstructed images of the graphs optimized on $S = \{0,1,2\}$.
    }
    \label{Figure: Histograms Particle Surface Area Distribution}

\end{figure*}

The RSA algorithm just places the original particles at new positions, such that the resulting images preserve the particle volume and surface area distributions exactly. To assess the fit after topological optimization and reconstruction, we computed histograms for the particle volume and particle surface area distributions of the real image, as well as the mean distributions of the $16$ RSA images, their reconstructions and the images optimized with $S = \{0,1,2\}$. To improve visibility, we excluded the largest $0.5\%$ of particles across the four images. The resulting histograms are shown in Figure~\ref{Figure: Histograms Particle Volume Distribution} and Figure~\ref{Figure: Histograms Particle Surface Area Distribution}. 

As expected, the particle volume and surface area distributions of the real and RSA images (blue) are nearly identical. The reconstructed and optimized images (green) also have distributions that are overall similar to the real distribution. Nevertheless, several differences can be observed across all four images. In the particle volume distribution, the volume of smaller particles tends to decrease further during reconstruction, whereas the frequencies of the largest particles increase. A likely explanation is that large particles might have large concave areas, which are filled by the convex hull during reconstructions. The particle surface area is shifted slightly towards smaller values. This is expected, since the convex hull produces locally flat boundaries, whereas the boundaries in the real images are more irregular. Overall, the reconstruction introduces only minor changes to both distributions while preserving their characteristics.

\begin{table}[t]
\centering
\textbf{Geometric Tortuosity}
\vspace{.2em}

{\small
\begin{tabular}{lcccc}
\toprule
Image Type & RC$1$ & RC$2$ & RC$3$ & RC$4$ \\
\midrule

Real Image & $1.157$ & $1.129$ & $1.146$ & $1.144$ \\
RSA & $1.652$ & $1.352$ & $1.297$ & $1.475$ \\
Reconstruction & $1.366$ & $1.259$ & $1.207$ & $1.324$ \\
$S=\{0\}$ & $1.307$ & $1.225$ & $1.183$ & $1.280$ \\
$S=\{1\}$ & $1.219$ & $1.176$ & $\mathbf{1.147}$ & $1.206$ \\
$S=\{2\}$ & $1.354$ & $1.243$ & $1.197$ & $1.306$ \\
$S=\{0,1\}$ & $1.216$ & $1.177$ & $1.148$ & $1.204$ \\
$S=\{0,2\}$ & $1.302$ & $1.217$ & $1.177$ & $1.269$ \\
$S=\{1,2\}$ & $1.217$ & $\mathbf{1.172}$ & $1.144$ & $1.201$ \\
$S=\{0,1,2\}$ & $\mathbf{1.214}$ & $1.173$ & $1.144$ & $\mathbf{1.199}$ \\
\bottomrule
\end{tabular}
}

\caption{The geometric tortuosity of the real image and the mean geometric tortuosities for the different synthetic images. Results closest to the geometric tortuosity of the real image are highlighted in bold.}
\label{Table: Geometric Tortuosity}
\end{table}

Finally, we study the geometric tortuosity of the real and the synthetic images. The geometric tortuosity is based on the geodesic distance between two foreground pixels, i.e., the length of the shortest path connecting two pixels while not leaving the foreground. The geometric tortuosity between the two pixels is then defined as the ratio of their geodesic distance and their Euclidean distance. For more details see Soille~\cite{Soille2004MorphologicalImageAnalysis}. To compute the geometric tortuosity of an image in a certain direction, one usually computes all geometric distances of border foreground pixels to the opposite image border and then averages the corresponding tortuosities.

We report the geometric tortuosity in the height direction of the scanning container. To avoid boundary effects, we computed the geodesic distances between the image slices $300$ and $1300$. The results are shown in Table~\ref{Table: Geometric Tortuosity}. We observe a positive domain gap from the RSA images to their reconstructions. This can be explained by two reasons. First, as indicated by the lower $H_0$ values in Table~\ref{Table: 1-Wasserstein distance synthetic to real image (normalized)}, the reconstructed images exhibit a higher connectivity than the RSA images. Consequently, they contain more possible paths, which reduces the geometric tortuosity. Second, the reconstructed convex particles allow for shorter paths than the original particles that may contain concavities. Since all the synthetic images exhibit a higher geometric tortuosity than the corresponding real images, both effects move the tortuosity in the desired direction. Furthermore, we observe that the optimization decreases the tortuosity further. Relative to the tortuosities of the reconstructed images, the best observed tortuosities reduce the deviation from the tortuosity of their respective real image by $70\%$ to $80\%$ for three of the four images and by $99\%$ for RC$3$. Comparing the tortuosities for different optimization sets $S$ reveals that the improvement is mainly driven by the optimization of the $1$-dimensional topology, while the best performance is achieved when optimizing over all three dimensions.

These results show that applying our optimization to synthetic particle systems yields consistent improvements in topological and geometric metrics, while only slightly affecting the particle geometry.
\subsection{Elasticity Analysis of Concrete Particle Systems}
\label{Subsection: Elasticity Analysis of Concrete Particle Systems}
The mechanical behavior of particle systems under external load is crucial to many applications. One important measure of this relationship is the Young's modulus~\cite{Callister_Rethwisch_2018_materials_intro}
\begin{equation}
    E = \frac{\sigma}{\epsilon}
\end{equation}
which measures the proportionality between applied stress $\sigma$ and the resulting strain in the particle system $\epsilon$. A higher Young's modulus indicates smaller deformation under the same stress and thus a stiffer system. Young's modulus depends on multiple properties of the particle system including its topology, volume fraction, particle shape, and contact areas.

In this section, we analyze the Young's moduli of the real and synthetic recycled concrete data in the coordinate directions. The physical sample itself is not a homogeneous system. When filling the scanning container, both the sedimentation process and the cylindrical boundary influence the particle system. We observe that the local volume fraction depends on the position in the cylinder. The local volume fraction generally increases when moving from the boundary towards the center of the cylinder. The reconstruction process amplifies this variation in volume fraction by reducing particle size at the boundary. Missing information outside the cylindrical sample causes no points to be sampled as edge features between the particle boundary and the cylinder wall, leading to smaller reconstructed particle sizes at the boundary. In contrast, the degree of particle graph nodes (i.e.\ the number of edges a node contributes to) is usually higher in the center, resulting in more points used in the convex hull computation and an increase in the reconstructed particle sizes.

For the computation of Young's modulus, we extract cuboidal crops from inside the cylinder. This simplifies the computation compared to simulating the stress on the cylinder boundaries. When extracting cuboidal crops from different real or synthetic images, it is necessary to approximately match the volume fractions to ensure comparability.

\begin{table}[t]
\centering
\textbf{Young's modulus}
\vspace{.2em}

{\small
\begin{tabular}{lcccc}
\toprule
Image & Vol.\ Frac.\ & $E_x$ & $E_y$ & $E_z$ \\
\midrule

RC$1$ & $0.528$ & $3.957$ & $3.577$ & $3.455$ \\
Reconstruction & $0.534$ & $3.682$ & $4.160$ & $3.895$ \\
RSA & $0.530$ & $1.506$ & $2.070$ & $1.093$ \\
Optimized & $0.549$ & $3.585$ & $2.544$ & $2.902$ \\
\bottomrule
\end{tabular}
}

\caption{The volume fraction and Young's moduli in $z$-, $y$- and $x$-direction of crops from RC$1$, its reconstruction, a synthetic RSA image, and an optimized reconstruction of the RSA image. The Young's moduli values are given in GPa.}
\label{Table: Young's modulus}
\end{table}

In the following, we use crops of size $350 \times 350 \times 1200$ voxels from four images: RC$1$, its reconstruction, a synthetic image generated using RSA, and its optimized reconstruction using $S = \{0,1,2\}$ with the same hyperparameters as in Section~\ref{Subsection: Synthetic Images of Concrete}. The Young's moduli are computed using GeoDict~\cite{geodict2026}. The results are shown in Table~\ref{Table: Young's modulus}. 

The reconstruction of the real image shows similar Young's moduli to the real image with variations both above and below. As expected, we observe the lowest values for the initial synthetic image generated using RSA. Since this generation method ignores gravity in the $z$-direction as well as particle contacts, the system is less stable, which results in a higher strain $\epsilon$ and lower Young's moduli. The optimized reconstruction increases the Young's moduli in all directions compared to the RSA image, but values remain below those of RC$1$. This means the topological optimization decreases the difference in mechanical behavior between the initial synthetic data and the real data. 

This can be explained by multiple factors. First, the higher volume fraction which is observed both in the crop and in the entire cylinder. Second, the topological optimization, particularly through the $0$- and $1$-dimensional topology, increases the number of particle contacts, making the system stiffer. It should be noted that besides the number of particle contacts also the contact area, which is not modeled during the topological optimization, influences the mechanical behavior. Nevertheless, these results confirm that postprocessing the initial synthetic data with our proposed topological optimization improves the mechanical behavior of synthetic particle systems.
\subsection{Application to Leaching in Active Protective Coatings}
\label{Subsection: Application to Leaching in Active Protective Coatings}

Active protective coatings are widely used to protect metallic components exposed to corrosive environments. In addition to providing a physical barrier between the corrosive environment and the metal, they have a second layer loaded with corrosion-inhibiting particles, which dissolve in the presence of water or other solvents. In the case of damage to the coating, they leach out of the coating and form a patch over the exposed surface~\cite{Visser_al2016AerospaceCoatings, VISSER_al2018importanceCoatings}.

The leaching process is complex and has attracted increasing interest due to efforts to replace the standard species for corrosion protection. Designing alternative active protective coatings is a complex and time consuming task. Modeling the inhibitor particles~\cite{Schladitz_al2024particlemodelscoating} and simulating the leaching process~\cite{Bonetti_al2024LeachingSimulation, Zaninovic2026LeachingSimulation} can accelerate the development of new active protective coatings.

The leaching process depends on many properties of the corrosion-inhibiting particle system such as the size and shape distribution of the particles as well as the inter particle structure. The currently used method to generate the synthetic data of corrosion-inhibiting particles is described in Schladitz et al.~\cite{Schladitz_al2024particlemodelscoating}. It primarily focuses on matching the size and shape distribution of the synthetic and real particles and then places them inside the observation window using RSA. However, the connectivity of the particles is a determining factor of the leaching process, which is currently not respected in the synthetic data. 

In the following, we compare the simulated leaching processes of a real corrosion-inhibitor particle system of lithium sulfate segmented from a CT scan, an initial synthetic inhibitor particle system matching the size and shape distribution of the real system, and a connected synthetic particle system obtained by postprocessing the initial synthetic particle system with our method. The leaching simulations were done on a $330 \times 330 \times 330$ crop of the complete image corresponding to a thickness of $15\si{\micro\meter}$.

In order to connect the synthetic particle system, we computed the particle graph on that crop of the synthetic image and then applied the topological optimization with $S = \{ 0 \}$, $\gamma=0$ and the other hyperparameters described in Appendix~\ref{Appendix: Experiment Details}. As the target diagram for the optimization we used an empty persistence diagram containing only the diagonal. The resulting image contains two connected components where the larger one makes up around $99.8\%$ of the total particle volume.

\begin{figure*}[t]
    \centering
    \textbf{Leaching Simulation Results of Lithium Sulfate Particles}

    \subfloat[Real Particle System]{
            \centering
            \includegraphics[width=0.32\textwidth]{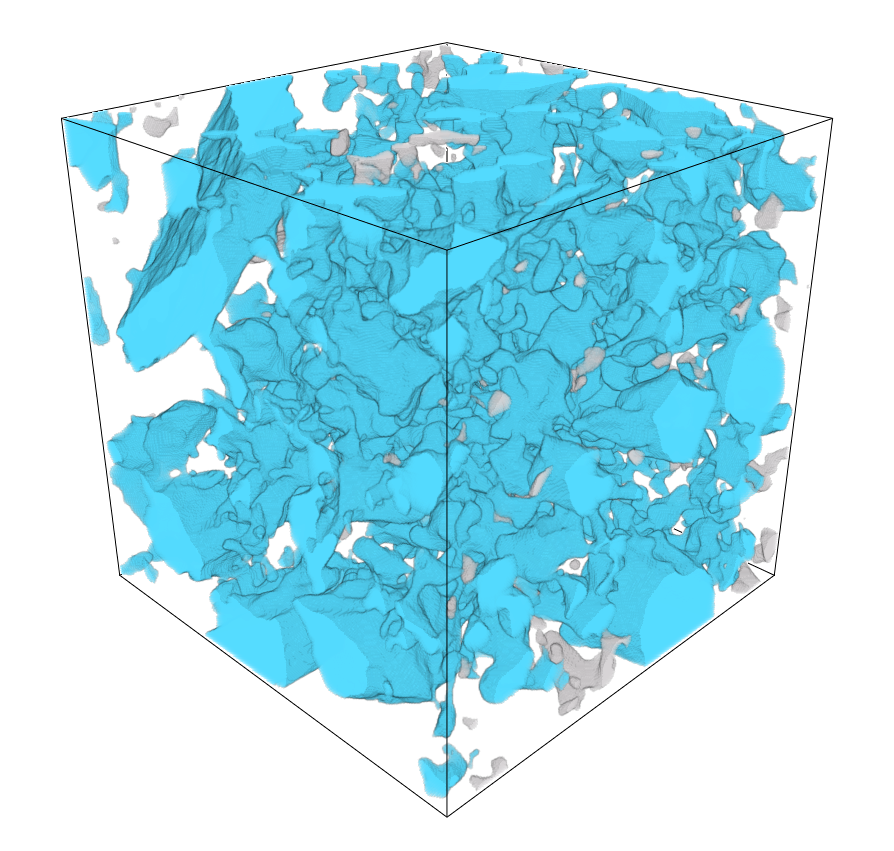}
    }
    \hfill    
    \subfloat[Initial Synthetic Particle System]{
            \centering
            \includegraphics[width=0.32\textwidth]{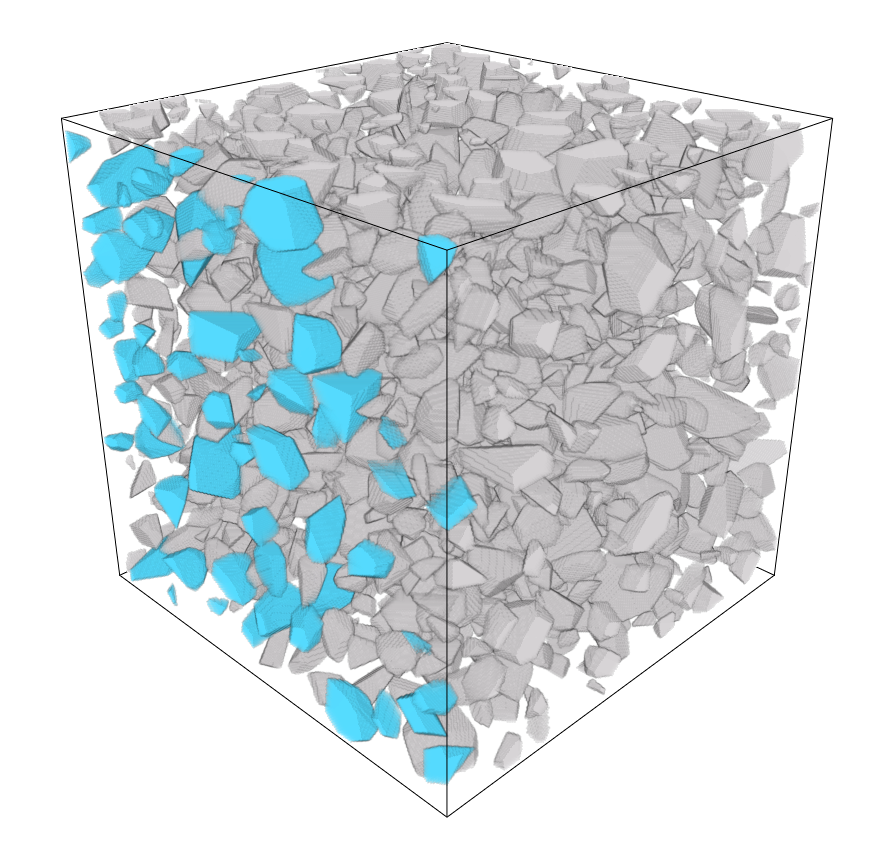}
    }
    \hfill
    \subfloat[Connected Synthetic Particle System]{
            \centering
            \includegraphics[width=0.32\textwidth]{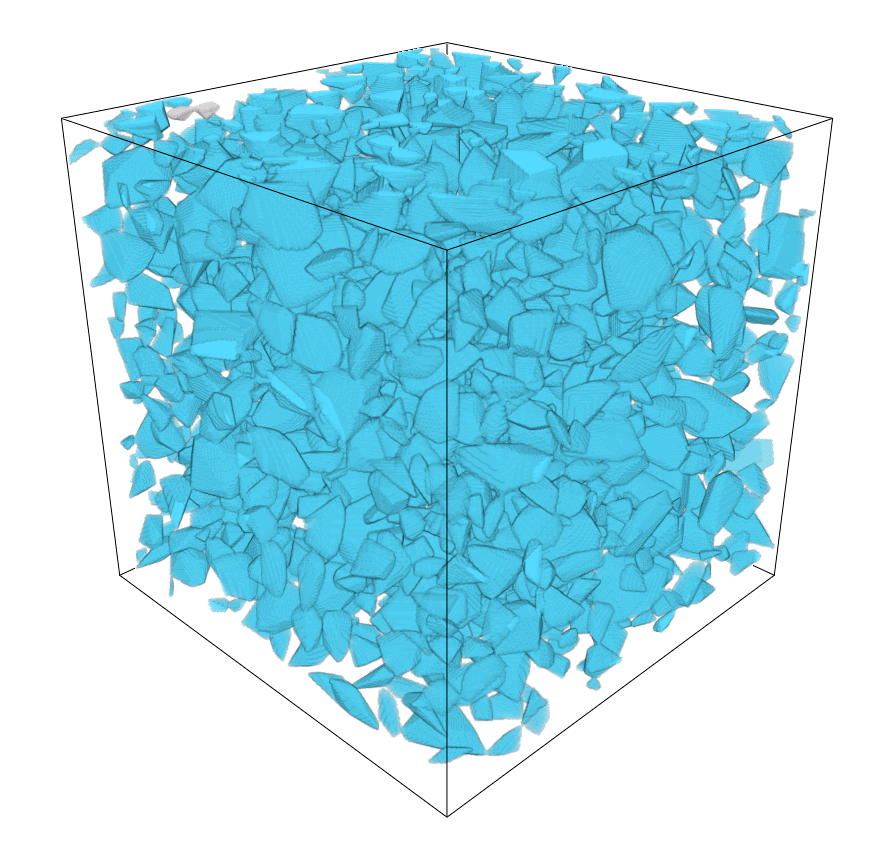}
    }

    \caption{Side-by-side comparison of the leaching simulation for (a) a real lithium sulfate corrosion-inhibiting particle system, (b) an initial synthetic particle system with fitted size and shape distribution, and (c) the with topological optimization postprocessed and connected synthetic particle systems. Particles dissolved by the leaching process are blue, remaining particles gray.}
    \label{Figure: Side-to-side comparison leaching simuluations}
\end{figure*}

The leaching simulation is carried out according to Zaninovi{\'c} et al.~\cite{Zaninovic2026LeachingSimulation}. Water comes in through one side of the observation window, representing a damaged region of the coating. Particles exposed at this boundary dissolve, allowing water to flow through the pores created by their dissolution and reach additional particles, which are then exposed to the leaching process. The simulation results are shown in Figure~\ref{Figure: Side-to-side comparison leaching simuluations}. We see that the initial synthetic particle system (b) is unsuitable for a leaching simulation in its current state, since its connectivity is too low. Consequently, only particles at the boundary being in contact with water are dissolved and the leaching process almost immediately stops. Postprocessing the synthetic particle system (c) allows the leaching process to propagate through the entire particle system and dissolve the particles in a manner similar to the real particle system (a).

In \cite{Zaninovic2026LeachingSimulation}, the authors overcame the unconnectedness of the initial synthetic particle system by applying a morphological closure, which created artificial tunnels between the particles. While such tunnels are not present in the real data prior to the exposure of water, Zaninovi{\'c} et al.~\cite{ZANINOVIC2026113631} showed in their in-situ experiment that similar tunnels form during the leaching process. While the morphological closure is primarily beneficial for creating an initial connectivity for the leaching simulation, it also simulates the tunnels which form during the leaching process in the real data. Our model should not be considered as an alternative to creating such tunnels in the synthetic data. It rather provides a more realistic initial synthetic particle system by increasing the number of particle contacts closer to that of the real system. This removes the need to introduce the tunnels solely to establish the initial connectivity, providing greater flexibility in how the tunnel formation during leaching is modeled. For example, the number of tunnels could be decreased by reducing the radius of the morphological closure or alternative approaches for the tunnel appearance could be tested. One possibility would be to let tunnels emerge dynamically during the leaching simulation as observed during the in-situ experiment. The effects of changing the synthetic particle systems to ones generated with our method have to be tested more extensively in the future for different particle systems and different particle species besides lithium sulfate such as barium sulfate, calcium sulfate, and zinc phosphate.

\section{Discussion}
\label{Section: Discussion}
This paper proposed an optimization scheme that can be applied to arbitrarily generated synthetic particle systems to improve topological and physical characteristics. To this end, we introduced a novel graph representation of the particle system, which leverages explicit particle information and offers computational advantages.

Our validation shows that our method consistently improves the topological quality and the geometric tortuosity of the generated images while preserving other important characteristics of the particle systems, including the number of particles and the distributions of particle volume and surface area. Furthermore, we have shown the applicability of our method to leaching simulations in active protective coatings. In this application, particle connectivity, represented by $0$-dimensional topological features, is the driving mechanism. As this is well captured by our approach, more realistic leaching simulations are obtained after postprocessing the initial particle configuration.

When looking at the $1$- and $2$-dimensional topological characteristics, the topological quality of the generated particles improves consistently, but there remains considerable potential for further improvement. The main limitation here is that the graph and image topology are not equivalent. As discussed in Section~\ref{Subsection: Relation between Image and Graph Topology}, the graph topology captures a lot of the topological features of the image correctly, but some features present in the image are not represented in the graph. This is mainly due to the reduction of information on particle shape, which is reconstructed as a convex hull. Hence, topological features appearing along the curved surface of neighboring particles cannot be reproduced. It should be noted that it is not just our method, but rather all methods that work with convex particles that have this limitation. For computational reasons, convexity is actually a pretty common assumption, see \cite{Schladitz_al2024particlemodelscoating, Escoda, Losch,  Zhang_2014}. As topological features introduced by non-convex particle surfaces are expected to be small, the impact on materials properties predictions might be negligible. However, this should be carefully checked in practice. Another weakness of the reconstruction method is the change in local volume fraction. While the real systems themselves are not homogeneous, the reconstruction amplifies the difference in the local volume fraction between the center of the cylinder and its boundary.

Narrowing the above mentioned domain gap introduced during the reconstruction through the graph representation is a promising direction for future work. In particular, replacing the convex hull used for the particle reconstruction from a set of points $X_v$ with a method capable of modeling concavities in the particle boundaries might likely improve the fit of the reconstruction and its local volume fraction. One possible alternative is the use of spherical harmonics conditioned on $X_v$ as boundary conditions~\cite{Feinhauer_al2015ParticleModellingSphericalHarmonics}. This approach introduces a probabilistic component into the reconstruction, whose influence, similar to the addition of simplex noise, would need to be investigated further. Another possible approach is the use of alpha shapes~\cite{Edelsbrunner_Mucke1994AlphaShapes} instead of the convex hull. However, the density of boundary points in $X_v$ is insufficient to reliably reconstruct particle boundaries using alpha shapes. While increasing the number of sampled boundary points per particle is straightforward, doing so does not allow for the same graph representation. The logical representation of the sampled points would be a marked point cloud and, as previously discussed, this would render the topological optimization computationally infeasible. Therefore, we expect that developing a method capable of exploiting a larger number of boundary points in combination with alpha shapes, while maintaining computational efficiency, could further improve the observed results.

Another direction for future work is computational efficiency of the method. While our proposed graph representation provides huge computational benefits, the runtime scales with the number of particles and the dimension up to which we compute persistence diagrams. Iteratively computing persistent homology up to dimension $2$ on graphs with more than $10{,}000$ nodes becomes computationally prohibitive. A promising direction for reducing this cost is the use of fast persistence updates based on computations obtained from previous iterations. Such approaches have been explored in various settings~\cite{Cohen-Steiner_al2006FastPersistenceUpdates,Luo_and_Nelseon2024FastPersistenceUpdates} and may also be applicable here. 

\section*{Data Availability}

The data generated for this work is published on Zenodo (doi: \url{https://doi.org/10.5281/zenodo.21868886}).

\section*{Code Availability}

The source code is available alongside the data on Zenodo as well as on GitHub (\href{https://github.com/martinmemmesheimer/fitting-topology-synthetic-particle-systems}{https://github.com/martinmemmesheimer/\allowbreak fitting-topology-synthetic-particle-systems})
\section*{Acknowledgements}
We thank Fraunhofer ITWM for providing the $\mu$CT images of recycled concrete, Hannes Grimm-Strele for the computation of the Young's moduli, and Jelena Zaninovi{\'c} for the simulation of the leaching process.

{
\sloppy
\hbadness=2200

\bibliographystyle{elsarticle-num}
\bibliography{references}

\begin{thebibliography}{10}
\expandafter\ifx\csname url\endcsname\relax
  \def\url#1{\texttt{#1}}\fi
\expandafter\ifx\csname urlprefix\endcsname\relax\def\urlprefix{URL }\fi
\expandafter\ifx\csname href\endcsname\relax
  \def\href#1#2{#2} \def\path#1{#1}\fi

\bibitem{Farhang_al2017GranularMaterialsDecade}
F.~Radjai, J.-N. Roux, A.~Daouadji, Modeling granular materials: Century-long research across scales, Journal of Engineering Mechanics 143~(4) (2017) 04017002.
\newblock \href {https://doi.org/10.1061/(ASCE)EM.1943-7889.0001196} {\path{doi:10.1061/(ASCE)EM.1943-7889.0001196}}.

\bibitem{Erdogan_al2006EngineeringApplication}
S.~Erdogan, P.~Quiroga, D.~Fowler, H.~Saleh, R.~Livingston, E.~Garboczi, P.~Ketcham, J.~Hagedorn, S.~Satterfield, Three-dimensional shape analysis of coarse aggregates: New techniques for and preliminary results on several different coarse aggregates and reference rocks, Cement and Concrete Research 36~(9) (2006) 1619--1627.
\newblock \href {https://doi.org/10.1016/j.cemconres.2006.04.003} {\path{doi:10.1016/j.cemconres.2006.04.003}}.

\bibitem{Fonseca_al2012EngineeringApplication}
J.~Fonseca, C.~O’Sullivan, M.~Coop, P.~Lee, Non-invasive characterization of particle morphology of natural sands, Soils and Foundations 52~(4) (2012) 712--722.
\newblock \href {https://doi.org/10.1016/j.sandf.2012.07.011} {\path{doi:10.1016/j.sandf.2012.07.011}}.

\bibitem{Garboczi2002EngineeringApplication}
E.~Garboczi, Three-dimensional mathematical analysis of particle shape using x-ray tomography and spherical harmonics: Application to aggregates used in concrete, Cement and Concrete Research 32~(10) (2002) 1621--1638.
\newblock \href {https://doi.org/10.1016/S0008-8846(02)00836-0} {\path{doi:10.1016/S0008-8846(02)00836-0}}.

\bibitem{Liu_al2020ConnectivityEngineeringApplication}
J.~Liu, W.~Zhou, G.~Ma, S.~Yang, X.~Chang, Strong contacts, connectivity and fabric anisotropy in granular materials: A 3d perspective, Powder Technology 366 (2020) 747--760.
\newblock \href {https://doi.org/https://doi.org/10.1016/j.powtec.2020.03.018} {\path{doi:https://doi.org/10.1016/j.powtec.2020.03.018}}.

\bibitem{Papadopoulos_al2018TopologyEngineeringApplication}
L.~Papadopoulos, M.~A. Porter, K.~E. Daniels, D.~S. Bassett, Network analysis of particles and grains, Journal of Complex Networks 6~(4) (2018) 485--565.
\newblock \href {https://doi.org/10.1093/comnet/cny005} {\path{doi:10.1093/comnet/cny005}}.

\bibitem{Walker_Tordesillas2010TopologyEngineeringApplication}
D.~M. Walker, A.~Tordesillas, Topological evolution in dense granular materials: A complex networks perspective, International Journal of Solids and Structures 47~(5) (2010) 624--639.
\newblock \href {https://doi.org/https://doi.org/10.1016/j.ijsolstr.2009.10.025} {\path{doi:https://doi.org/10.1016/j.ijsolstr.2009.10.025}}.

\bibitem{Lin_al2022MacroscopicBehaviour}
M.~Lin, W.~Zhou, J.~Liu, G.~Ma, X.~Cao, A topological view on microscopic structural evolution for granular material under loading and unloading path, Computers and Geotechnics 141 (2022) 104530.
\newblock \href {https://doi.org/https://doi.org/10.1016/j.compgeo.2021.104530} {\path{doi:https://doi.org/10.1016/j.compgeo.2021.104530}}.

\bibitem{ZhaoandZhou2017MacroscopicBehaviour}
S.~Zhao, X.~Zhou, Effects of particle asphericity on the macro- and micro-mechanical behaviors of granular assemblies, Granular Matter 19~(2) (2017) 38.
\newblock \href {https://doi.org/10.1007/s10035-017-0725-6} {\path{doi:10.1007/s10035-017-0725-6}}.

\bibitem{ZANINOVIC2026113631}
J.~Zaninović, T.~Nogatz, K.~Schladitz, S.~Flenner, I.~Greving, P.~Klein, N.~Konchakova, C.~Redenbach, P.~Visser, Leaching in active protective coatings observed in-situ by nano-{CT} using synchrotron radiation, Corrosion Science 263 (2026) 113631.
\newblock \href {https://doi.org/https://doi.org/10.1016/j.corsci.2026.113631} {\path{doi:https://doi.org/10.1016/j.corsci.2026.113631}}.

\bibitem{chiu_al2013stochasticgeometry}
S.~N. Chiu, D.~Stoyan, W.~S. Kendall, J.~Mecke, Stochastic geometry and its applications, John Wiley \& Sons, 2013.

\bibitem{You_al2019SyntheticParticleComposites}
H.~You, Y.~Kim, G.~J. Yun, Computationally fast morphological descriptor-based microstructure reconstruction algorithms for particulate composites, Composites Science and Technology 182 (2019) 107746.
\newblock \href {https://doi.org/https://doi.org/10.1016/j.compscitech.2019.107746} {\path{doi:https://doi.org/10.1016/j.compscitech.2019.107746}}.

\bibitem{Feinhauer_al2015ParticleModellingSphericalHarmonics}
J.~Feinauer, T.~Brereton, A.~Spettl, M.~Weber, I.~Manke, V.~Schmidt, Stochastic 3d modeling of the microstructure of lithium-ion battery anodes via gaussian random fields on the sphere, Computational Materials Science 109 (2015) 137--146.
\newblock \href {https://doi.org/https://doi.org/10.1016/j.commatsci.2015.06.025} {\path{doi:https://doi.org/10.1016/j.commatsci.2015.06.025}}.

\bibitem{Wang_al2025SyntheticParticlePack}
B.~Wang, A.~M. Kingston, P.~D. Lösel, W.~Creemers, Synthetic particle pack generation for augmentation and testing in geological tomographic segmentation, Tomography of Materials and Structures 9 (2025) 100072.
\newblock \href {https://doi.org/https://doi.org/10.1016/j.tmater.2025.100072} {\path{doi:https://doi.org/10.1016/j.tmater.2025.100072}}.

\bibitem{Gupta_al2025TopoDiffusionNet}
S.~Gupta, D.~Samaras, C.~Chen, \href{https://proceedings.iclr.cc/paper_files/paper/2025/file/4e889e581d4a6f2be932c6f65e7792a8-Paper-Conference.pdf}{Topodiffusionnet: A topology-aware diffusion model}, in: Y.~Yue, A.~Garg, N.~Peng, F.~Sha, R.~Yu (Eds.), International Conference on Learning Representations, Vol. 2025, 2025, pp. 31699--31713.
\urlprefix\url{https://proceedings.iclr.cc/paper_files/paper/2025/file/4e889e581d4a6f2be932c6f65e7792a8-Paper-Conference.pdf}

\bibitem{Liu_al2025TopoLiDM}
J.~Liu, Z.~Huang, M.~Liu, T.~Deng, F.~Nex, H.~Cheng, H.~Wang, Topolidm: Topology-aware lidar diffusion models for interpretable and realistic lidar point cloud generation, in: C.~Laugier, A.~Renzaglia, N.~Atanasov, S.~Birchfield, G.~Cielniak, L.~\{De Mattos\}, L.~Fiorini, P.~Giguere, K.~Hashimoto, J.~Ibanez-Guzman, T.~Kamegawa, J.~Lee, G.~Loianno, K.~Luck, H.~Maruyama, P.~Martinet, H.~Moradi, U.~Nunes, J.~Pettre, A.~Pretto, T.~Ranzani, A.~Ronnau, S.~Rossi, E.~Rouse, F.~Ruggiero, O.~Simonin, D.~Wang, M.~Yang, E.~Yoshida, H.~Zhao (Eds.), 2025 IEEE/RSJ International Conference on Intelligent Robots and Systems (IROS), IEEE, 2025, pp. 8180--8186.
\newblock \href {https://doi.org/10.1109/IROS60139.2025.11247363} {\path{doi:10.1109/IROS60139.2025.11247363}}.

\bibitem{Wang_al2020TopoGAN}
F.~Wang, H.~Liu, D.~Samaras, C.~Chen, Topogan: A topology-aware generative adversarial network, in: A.~Vedaldi, H.~Bischof, T.~Brox, J.-M. Frahm (Eds.), Computer Vision -- ECCV 2020, Springer International Publishing, Cham, 2020, pp. 118--136.

\bibitem{Poulenard_al2018SurfaceMatching}
A.~Poulenard, P.~Skraba, M.~Ovsjanikov, Topological function optimization for continuous shape matching, Computer Graphics Forum 37~(5) (2018) 13--25.
\newblock \href {https://doi.org/10.1111/cgf.13487} {\path{doi:10.1111/cgf.13487}}.

\bibitem{BruelGabrielson_al2020SurfaceReconstructionForPointClouds}
R.~Brüel-Gabrielsson, V.~Ganapathi-Subramanian, P.~Skraba, L.~J. Guibas, Topology-aware surface reconstruction for point clouds, Computer Graphics Forum 39~(5) (2020) 197--207.
\newblock \href {https://doi.org/10.1111/cgf.14079} {\path{doi:10.1111/cgf.14079}}.

\bibitem{Kissi_al2025TopologicalSimplification}
M.~Kissi, M.~Pont, J.~A. Levine, J.~Tierny, A practical solver for scalar data topological simplification, IEEE Transactions on Visualization and Computer Graphics 31~(1) (2025) 97--107.
\newblock \href {https://doi.org/10.1109/TVCG.2024.3456345} {\path{doi:10.1109/TVCG.2024.3456345}}.

\bibitem{Carriere_al2021OptimizingPHbasedFunctions}
M.~Carriere, F.~Chazal, M.~Glisse, Y.~Ike, H.~Kannan, Y.~Umeda, \href{https://proceedings.mlr.press/v139/carriere21a.html}{Optimizing persistent homology based functions}, in: M.~Meila, T.~Zhang (Eds.), Proceedings of the 38th International Conference on Machine Learning, Vol. 139 of Proceedings of Machine Learning Research, PMLR, 2021, pp. 1294--1303.
\urlprefix\url{https://proceedings.mlr.press/v139/carriere21a.html}

\bibitem{Edelsbrunner_al2002Intro_PH}
Edelsbrunner, Letscher, Zomorodian, Topological persistence and simplification, Discrete \& computational geometry 28~(4) (2002) 511--533.
\newblock \href {https://doi.org/10.1090/mbk/069} {\path{doi:10.1090/mbk/069}}.

\bibitem{Zomorodian_Carlson2005Intro_PH}
A.~Zomorodian, G.~Carlsson, Computing persistent homology, Discrete \& Computational Geometry 33~(2) (2005) 249--274.
\newblock \href {https://doi.org/10.1007/s00454-004-1146-y} {\path{doi:10.1007/s00454-004-1146-y}}.

\bibitem{Freudenthal1942Triangulation}
H.~Freudenthal, \href{http://www.jstor.org/stable/1968813}{Simplizialzerlegungen von beschränkter {Flachheit}}, Annals of Mathematics 43~(3) (1942) 580--582.
\urlprefix\url{http://www.jstor.org/stable/1968813}

\bibitem{Edelsbrunner_Harer2010IntroComputationalTopology}
H.~Edelsbrunner, J.~Harer, Computational topology: an introduction, American Mathematical Soc., 2010.

\bibitem{Kaczynski_al2004IntroCubicalHomology}
T.~Kaczynski, K.~Mischaikow, M.~Mrozek, Cubical Homology, Springer New York, New York, NY, 2004.
\newblock \href {https://doi.org/10.1007/0-387-21597-2_2} {\path{doi:10.1007/0-387-21597-2_2}}.

\bibitem{Kramar_al2014ForceNetworksPH}
M.~Kram\'ar, A.~Goullet, L.~Kondic, K.~Mischaikow, Evolution of force networks in dense particulate media, Phys. Rev. E 90 (2014) 052203.
\newblock \href {https://doi.org/10.1103/PhysRevE.90.052203} {\path{doi:10.1103/PhysRevE.90.052203}}.

\bibitem{Mei_al2022ProximityGraph}
J.~Mei, G.~Ma, Q.~Wang, T.~Wu, W.~Zhou, Micro- and macroscopic aspects of the intermittent behaviors of granular materials related by graph neural network, International Journal of Solids and Structures 251 (2022) 111763.
\newblock \href {https://doi.org/https://doi.org/10.1016/j.ijsolstr.2022.111763} {\path{doi:https://doi.org/10.1016/j.ijsolstr.2022.111763}}.

\bibitem{Duran_al2010ProximityGraph}
O.~Durán, N.~Kruyt, S.~Luding, Micro-mechanical analysis of deformation characteristics of three-dimensional granular materials, International Journal of Solids and Structures 47~(17) (2010) 2234--2245.
\newblock \href {https://doi.org/https://doi.org/10.1016/j.ijsolstr.2010.04.014} {\path{doi:https://doi.org/10.1016/j.ijsolstr.2010.04.014}}.

\bibitem{Bresenham1965LineAlgorithm}
J.~E. Bresenham, Algorithm for computer control of a digital plotter, IBM Systems Journal 4~(1) (1965) 25--30.
\newblock \href {https://doi.org/10.1147/sj.41.0025} {\path{doi:10.1147/sj.41.0025}}.

\bibitem{Trimesh2019}
M.~Dawson-Haggerty, J.~Bacher, A.~Brown-Bivins, C.~Forest, J.~O'Kane, \href{https://trimesh.org/}{trimesh: Load and analyze triangular meshes} (2019).
\urlprefix\url{https://trimesh.org/}

\bibitem{Ohser_Schladitz2009ImageProcessingBook}
J.~Ohser, K.~Schladitz, 3D Images of Materials Structures, John Wiley \& Sons, Ltd, 2009.
\newblock \href {https://doi.org/10.1002/9783527628308} {\path{doi:10.1002/9783527628308}}.

\bibitem{Gustavson2005Simplex_Noise}
S.~Gustavson, Simplex noise demystified (01 2005).

\bibitem{Perlin2001Simplex_noise}
K.~Perlin, Noise hardware. in real-time shading, SIGGRAPH Course Notes (2001).

\bibitem{carriere_al2026surveytopooptim}
M.~Carriere, Y.~Ike, T.~Lacombe, N.~Nishikawa, Persistence-based topological optimization: a survey, arXiv preprint arXiv:2603.24613 (2026).
\newblock \href {https://doi.org/10.48550/arXiv.2603.24613} {\path{doi:10.48550/arXiv.2603.24613}}.

\bibitem{CubicalRipser2020}
S.~Kaji, T.~Sudo, K.~Ahara, Cubical ripser: Software for computing persistent homology of image and volume data (2020).
\newblock \href {http://arxiv.org/abs/arXiv:2005.12692} {\path{arXiv:arXiv:2005.12692}}.

\bibitem{burella2021giottoph}
J.~B. Pérez, S.~Hauke, U.~Lupo, M.~Caorsi, A.~Dassatti, giotto-ph: A python library for high-performance computation of persistent homology of vietoris--rips filtrations (2021).
\newblock \href {http://arxiv.org/abs/2107.05412} {\path{arXiv:2107.05412}}.

\bibitem{Feldkamp_al1984ReconstructionCTImage}
L.~A. Feldkamp, L.~C. Davis, J.~W. Kress, Practical cone-beam algorithm, J. Opt. Soc. Am. A 1~(6) (1984) 612--619.
\newblock \href {https://doi.org/10.1364/JOSAA.1.000612} {\path{doi:10.1364/JOSAA.1.000612}}.

\bibitem{Otsu1979Thresholding}
N.~Otsu, A threshold selection method from gray-level histograms, IEEE Transactions on Systems, Man, and Cybernetics 9~(1) (1979) 62--66.
\newblock \href {https://doi.org/10.1109/TSMC.1979.4310076} {\path{doi:10.1109/TSMC.1979.4310076}}.

\bibitem{Vincent_and_Soille1991Watershed}
L.~Vincent, P.~Soille, Watersheds in digital spaces: an efficient algorithm based on immersion simulations, IEEE Transactions on Pattern Analysis and Machine Intelligence 13~(6) (1991) 583--598.
\newblock \href {https://doi.org/10.1109/34.87344} {\path{doi:10.1109/34.87344}}.

\bibitem{Burgmann_al2022DetailsOfData}
S.~Burgmann, M.~Godehardt, K.~Schladitz, W.~Breit, Separation of sand and gravel particles in 3d images using the adaptive h-extrema transform, Powder Technology 404 (2022) 117468.
\newblock \href {https://doi.org/10.1016/j.powtec.2022.117468} {\path{doi:10.1016/j.powtec.2022.117468}}.

\bibitem{Mohapatra_al2023dice_in_survey}
R.~K. Mohapatra, L.~Jolly, D.~C. Lyngdoh, G.~K. Mourya, I.~A. Changaai~Mangalote, S.~I. Alam, S.~P. Dakua, A comprehensive survey to study the utilities of image segmentation methods in clinical routine, Network Modeling Analysis in Health Informatics and Bioinformatics 13~(1) (2023) 2.
\newblock \href {https://doi.org/10.1007/s13721-023-00436-z} {\path{doi:10.1007/s13721-023-00436-z}}.

\bibitem{Kingma_and_Ba2015AdamOptimization}
D.~P. Kingma, J.~Ba, \href{https://mlanthology.org/iclr/2015/kingma2015iclr-adam/}{{Adam: A Method for Stochastic Optimization}}, in: Y.~Bengio, Y.~LeCun (Eds.), 3rd International Conference on Learning Representations, Conference Track Proceedings, 2015.
\urlprefix\url{https://mlanthology.org/iclr/2015/kingma2015iclr-adam/}

\bibitem{Soille2004MorphologicalImageAnalysis}
P.~Soille, Morphological Image Analysis: Principles and Applications, Springer Berlin Heidelberg, Berlin, Heidelberg, 2004.
\newblock \href {https://doi.org/10.1007/978-3-662-05088-0} {\path{doi:10.1007/978-3-662-05088-0}}.

\bibitem{Callister_Rethwisch_2018_materials_intro}
W.~D. Callister, D.~G. Rethwisch, Materials {Science} and {Engineering}: {An} {Introduction}, 10th Edition, Wiley, Hoboken, 2018, original Date: 2000.

\bibitem{geodict2026}
{Math2Market GmbH}, Geodict simulation software release 2026 (2026).
\newblock \href {https://doi.org/10.30423/release.geodict2026} {\path{doi:10.30423/release.geodict2026}}.

\bibitem{Visser_al2016AerospaceCoatings}
P.~Visser, H.~Terryn, J.~M.~C. Mol, Aerospace Coatings, Springer Netherlands, Dordrecht, 2016, pp. 315--372.
\newblock \href {https://doi.org/10.1007/978-94-017-7540-3_12} {\path{doi:10.1007/978-94-017-7540-3_12}}.

\bibitem{VISSER_al2018importanceCoatings}
P.~Visser, H.~Terryn, J.~Mol, On the importance of irreversibility of corrosion inhibitors for active coating protection of aa2024-t3, Corrosion Science 140 (2018) 272--285.
\newblock \href {https://doi.org/https://doi.org/10.1016/j.corsci.2018.05.037} {\path{doi:https://doi.org/10.1016/j.corsci.2018.05.037}}.

\bibitem{Schladitz_al2024particlemodelscoating}
K.~Schladitz, C.~Jung, S.~Flenner, M.~Godehardt, B.~Grevelhörster, I.~Greving, P.~Klein, N.~Konchakova, C.~Redenbach, P.~Visser, J.~Zaninović, Geometric modelling of corrosion inhibitor pigments in active protective coatings based on {SR-nano-CT} images, Progress in Organic Coatings 197 (2024) 108762.
\newblock \href {https://doi.org/10.1016/j.porgcoat.2024.108762} {\path{doi:10.1016/j.porgcoat.2024.108762}}.

\bibitem{Bonetti_al2024LeachingSimulation}
E.~Bonetti, P.~Visser, F.~R. Siperstein, S.~R. Gibbon, 5 - a cellular automaton paint to study the leaching of active corrosion inhibitors from primers, in: D.~Féron, D.~{di Caprio} (Eds.), Corrosion Modelling with Cellular Automata, Vol.~71 of European Federation of Corrosion (EFC) Series, Woodhead Publishing, 2024, pp. 89--107.
\newblock \href {https://doi.org/https://doi.org/10.1016/B978-0-443-13786-0.00002-2} {\path{doi:https://doi.org/10.1016/B978-0-443-13786-0.00002-2}}.

\bibitem{Zaninovic2026LeachingSimulation}
J.~Zaninovi{\'c}, S.~Osterroth, C.~Redenbach, et~al., Stochastic geometry models for inhibitor particles in active-reactive coatings and their impact on leaching simulations on the micro-scalePreprint, Version 1, Research Square (Sep. 2026).
\newblock \href {https://doi.org/10.21203/rs.3.rs-10898159/v1} {\path{doi:10.21203/rs.3.rs-10898159/v1}}.

\bibitem{Escoda}
J.~Escoda, D.~Jeulin, F.~Willot, C.~Toulemonde, Three-dimensional morphological modelling of concrete using multiscale poisson polyhedra, Journal of Microscopy 258~(1) (2015) 31--48.
\newblock \href {http://arxiv.org/abs/https://onlinelibrary.wiley.com/doi/pdf/10.1111/jmi.12213} {\path{arXiv:https://onlinelibrary.wiley.com/doi/pdf/10.1111/jmi.12213}}, \href {https://doi.org/https://doi.org/10.1111/jmi.12213} {\path{doi:https://doi.org/10.1111/jmi.12213}}.

\bibitem{Losch}
K.~Losch, S.~Schuff, F.~Balle, T.~Beck, C.~Redenbach, A stochastic microstructure model for particle reinforced aluminium matrix composites, Journal of Microscopy 273~(2) (2019) 115--126.
\newblock \href {http://arxiv.org/abs/https://onlinelibrary.wiley.com/doi/pdf/10.1111/jmi.12766} {\path{arXiv:https://onlinelibrary.wiley.com/doi/pdf/10.1111/jmi.12766}}, \href {https://doi.org/https://doi.org/10.1111/jmi.12766} {\path{doi:https://doi.org/10.1111/jmi.12766}}.

\bibitem{Zhang_2014}
X.~X. Zhang, Q.~Zhang, T.~Zangmeister, B.~L. Xiao, H.~Andrä, Z.~Y. Ma, A three-dimensional realistic microstructure model of particle-reinforced metal matrix composites, Modelling and Simulation in Materials Science and Engineering 22~(3) (2014) 035010.
\newblock \href {https://doi.org/10.1088/0965-0393/22/3/035010} {\path{doi:10.1088/0965-0393/22/3/035010}}.

\bibitem{Edelsbrunner_Mucke1994AlphaShapes}
H.~Edelsbrunner, E.~P. M\"{u}cke, Three-dimensional alpha shapes, ACM Trans. Graph. 13~(1) (1994) 43–72.
\newblock \href {https://doi.org/10.1145/174462.156635} {\path{doi:10.1145/174462.156635}}.

\bibitem{Cohen-Steiner_al2006FastPersistenceUpdates}
D.~Cohen-Steiner, H.~Edelsbrunner, D.~Morozov, Vines and vineyards by updating persistence in linear time, in: N.~Amenta, O.~Cheong (Eds.), Proceedings of the Twenty-Second Annual Symposium on Computational Geometry, SCG '06, Association for Computing Machinery, New York, NY, USA, 2006, pp. 119--126.
\newblock \href {https://doi.org/10.1145/1137856.1137877} {\path{doi:10.1145/1137856.1137877}}.

\bibitem{Luo_and_Nelseon2024FastPersistenceUpdates}
Y.~Luo, B.~J. Nelson, Accelerating iterated persistent homology computations with warm starts, Computational Geometry 120 (2024) 102089.
\newblock \href {https://doi.org/https://doi.org/10.1016/j.comgeo.2024.102089} {\path{doi:https://doi.org/10.1016/j.comgeo.2024.102089}}.

\bibitem{Divol_al2021geometryPDspace}
V.~Divol, T.~Lacombe, Understanding the topology and the geometry of the space of persistence diagrams via optimal partial transport, Journal of Applied and Computational Topology 5~(1) (2021) 1--53.
\newblock \href {https://doi.org/10.1007/s41468-020-00061-z} {\path{doi:10.1007/s41468-020-00061-z}}.

\bibitem{Chizat_al2018unbalancedOT}
L.~Chizat, G.~Peyré, B.~Schmitzer, F.-X. Vialard, Unbalanced optimal transport: Dynamic and kantorovich formulations, Journal of Functional Analysis 274~(11) (2018) 3090--3123.
\newblock \href {https://doi.org/10.1016/j.jfa.2018.03.008} {\path{doi:10.1016/j.jfa.2018.03.008}}.

\bibitem{Figali_Gigli2010unbalancedOT}
A.~Figalli, N.~Gigli, A new transportation distance between non-negative measures, with applications to gradients flows with dirichlet boundary conditions, Journal de Mathématiques Pures et Appliquées 94~(2) (2010) 107--130.
\newblock \href {https://doi.org/10.1016/j.matpur.2009.11.005} {\path{doi:10.1016/j.matpur.2009.11.005}}.

\bibitem{Kondratyev_al2016unbalancedOT}
S.~Kondratyev, L.~Monsaingeon, D.~Vorotnikov, {A new optimal transport distance on the space of finite Radon measures}, Advances in Differential Equations 21~(11/12) (2016) 1117 -- 1164.
\newblock \href {https://doi.org/10.57262/ade/1476369298} {\path{doi:10.57262/ade/1476369298}}.

\bibitem{lacombe_al2018optimaltransport}
T.~Lacombe, M.~Cuturi, S.~Oudot, \href{https://proceedings.neurips.cc/paper_files/paper/2018/file/b58f7d184743106a8a66028b7a28937c-Paper.pdf}{Large scale computation of means and clusters for persistence diagrams using optimal transport}, in: S.~Bengio, H.~Wallach, H.~Larochelle, K.~Grauman, N.~Cesa-Bianchi, R.~Garnett (Eds.), Advances in Neural Information Processing Systems, Vol.~31, Curran Associates, Inc., 2018.
\urlprefix\url{https://proceedings.neurips.cc/paper_files/paper/2018/file/b58f7d184743106a8a66028b7a28937c-Paper.pdf}

\bibitem{Cuturi2013regularizedoptimaltransport}
M.~Cuturi, \href{https://proceedings.neurips.cc/paper_files/paper/2013/file/af21d0c97db2e27e13572cbf59eb343d-Paper.pdf}{Sinkhorn distances: Lightspeed computation of optimal transport}, in: C.~Burges, L.~Bottou, M.~Welling, Z.~Ghahramani, K.~Weinberger (Eds.), Advances in Neural Information Processing Systems, Vol.~26, Curran Associates, Inc., 2013.
\urlprefix\url{https://proceedings.neurips.cc/paper_files/paper/2013/file/af21d0c97db2e27e13572cbf59eb343d-Paper.pdf}

\bibitem{Furat_al2021outershellmodel}
O.~Furat, L.~Petrich, D.~P. Finegan, D.~Diercks, F.~Usseglio-Viretta, K.~Smith, V.~Schmidt, Artificial generation of representative single li-ion electrode particle architectures from microscopy data, npj Computational Materials 7~(1) (2021) 105.
\newblock \href {https://doi.org/10.1038/s41524-021-00567-9} {\path{doi:10.1038/s41524-021-00567-9}}.

\bibitem{Paszke_al2019PytorchReference}
A.~Paszke, S.~Gross, F.~Massa, A.~Lerer, J.~Bradbury, G.~Chanan, T.~Killeen, Z.~Lin, N.~Gimelshein, L.~Antiga, A.~Desmaison, A.~Kopf, E.~Yang, Z.~DeVito, M.~Raison, A.~Tejani, S.~Chilamkurthy, B.~Steiner, L.~Fang, J.~Bai, S.~Chintala, \href{https://proceedings.neurips.cc/paper_files/paper/2019/file/bdbca288fee7f92f2bfa9f7012727740-Paper.pdf}{Pytorch: An imperative style, high-performance deep learning library}, in: H.~Wallach, H.~Larochelle, A.~Beygelzimer, F.~d\textquotesingle Alch\'{e}-Buc, E.~Fox, R.~Garnett (Eds.), Advances in Neural Information Processing Systems, Vol.~32, Curran Associates, Inc., 2019.
\urlprefix\url{https://proceedings.neurips.cc/paper_files/paper/2019/file/bdbca288fee7f92f2bfa9f7012727740-Paper.pdf}

\bibitem{Feydy_al2019GeomlossPackage}
J.~Feydy, T.~S\'{e}journ\'{e}, F.-X. Vialard, S.-i. Amari, A.~Trouve, G.~Peyr\'{e}, \href{https://proceedings.mlr.press/v89/feydy19a.html}{Interpolating between optimal transport and mmd using sinkhorn divergences}, in: K.~Chaudhuri, M.~Sugiyama (Eds.), Proceedings of the Twenty-Second International Conference on Artificial Intelligence and Statistics, Vol.~89 of Proceedings of Machine Learning Research, PMLR, 2019, pp. 2681--2690.
\urlprefix\url{https://proceedings.mlr.press/v89/feydy19a.html}

\end{thebibliography}
}

\appendix

\makeatletter
\patchcmd{\@seccntformat}{\csname the#1\endcsname}{Appendix \csname the#1\endcsname}{}{}
\makeatother

\counterwithin{equation}{section}
\renewcommand{\theequation}{\thesection.\arabic{equation}}

\counterwithin{figure}{section}
\renewcommand{\thefigure}{\thesection.\arabic{figure}}

\counterwithin{table}{section}
\renewcommand{\thetable}{\thesection.\arabic{table}}

\section{Entropic Regularization}
\label{Appendix: Entropic Regularization}

Optimal transport is closely related to Wasserstein distances and can be used to study the space of persistence diagrams, see Divol and Lacombe~\cite{Divol_al2021geometryPDspace}. It provides tools to compare measures by seeking an optimal transport plan which minimizes the cost of transporting the mass from one measure to another. The connection to persistence diagrams arises by representing a persistence diagram as a finite sum of Dirac measures located at its off-diagonal points. However, standard optimal transport theory assumes the two measures to have equal mass, which is generally not the case for persistence diagrams. An extension of the optimal transport problem to measures of different mass has been discussed by various authors~\cite{Chizat_al2018unbalancedOT,Figali_Gigli2010unbalancedOT,Kondratyev_al2016unbalancedOT}. Lacombe et al.~\cite{lacombe_al2018optimaltransport} show how the Wasserstein distance between persistence diagrams can be computed using the discrete optimal transport formulation. Then they use an entropically regularized version of this problem, introduced by Cuturi~\cite{Cuturi2013regularizedoptimaltransport}, to approximate the Wasserstein distance. This regularized version offers several advantages, as it is faster to compute, differentiable and provides more stable gradients. In the following we introduce the necessary background in optimal transport theory, following the notation in Lacombe et al.~\cite{lacombe_al2018optimaltransport}.

Consider two finite discrete measures $\mu = \sum_{i=1}^{n} a_i \delta_{x_i}$ and $\nu = \sum_{i=1}^{m} b_i \delta_{y_i}$ with positive weights $a_i, b_j \in \mathbb{R}_+$ and points from the upper half plane $x_i, y_j \in R_>^2$ for $i=1, \ldots,n$ and $j=1, \ldots, m$. The discrete optimal transport problem requires that the measures have equal mass, i.e.\ they must satisfy $\sum_{i=1}^{n} a_i = \sum_{j=1}^{m} b_j$. The aim is to compute the minimal transportation cost from $\mu$ to $\nu$ w.r.t.\ some cost matrix $C \in \mathbb{R}^{n \times m}$. A transportation plan is a matrix $P \in \mathbb{R}^{n \times m}$ whose rows and columns sum up to $a = (a_1, \ldots, a_n)^T$ and $b=(b_1, \ldots, b_m)^T$, respectively. The transportation polytope
\begin{equation}
    \Pi(a,b) \coloneqq \{P \in \mathbb{R}_+^{n \times m} \ | \ P\textbf{1}_m = a, P^T\textbf{1}_n = b \}.
\end{equation}
is the set of all transportation plans and the cost of a plan $P \in \Pi(a,b)$ is given by the Frobenius dot product between $P$ and $C$. The discrete optimal transport problem is then given by
\begin{equation}
    \label{OT_LP}
    \textbf{L}_C(\mu, \nu) \coloneqq \min_{P \in \Pi(a,b)} \langle P,C \rangle.
\end{equation}
Solving this linear program becomes intractable for large $n,m$. To address this, Cuturi~\cite{Cuturi2013regularizedoptimaltransport} proposed to add an entropic regularizer to the formulation. The regularization term is defined as $h(P) \coloneqq - \sum_{i,j} P_{ij} (\log P_{ij} - 1)$, and for $\gamma > 0$ the regularized problem is given by
\begin{equation}
    \label{Equation: Regularized discrete optimal transport}
    \textbf{L}_C^\gamma(\mu, \nu) \coloneqq \min_{P \in \Pi(a,b)} \left< P,C \right> - \gamma h(P).
\end{equation}
This formulation has several desirable properties. In particular, it can be solved significantly faster than the unregularized version using Sinkhorn divergences, it has a unique solution $P^{\gamma}$, and the cost $\langle P^{\gamma}, C \rangle$ of the optimal transport goes to $\textbf{L}_C(\mu, \nu)$ as $\gamma \rightarrow 0$. Furthermore, it is differentiable and provides stable gradients, which is most important for our use. By expressing the $2$-Wasserstein distance as a discrete optimal transport problem, we can use Eq.~\eqref{Equation: Regularized discrete optimal transport} in our loss term. 

A persistence diagram can be associated with a finite sum of Dirac measures $\sum_{i=1}^{n} a_i \delta_{x_i}$ where $x_i \in \mathbb{R}_>^2$ are the off-diagonal points in the diagram and the $a_i \in \mathbb{N}$ are their multiplicities. To maintain the behavior of the diagonal in this association, a virtual diagonal point $\{\Delta\}$ is added to the measure. Formally, this is done by a linear operator $R \colon \mu \mapsto |\mu| \delta_{\{\Delta\}}$, which assigns the mass of a measure to $\{\Delta\}$. Consider the associated measures of two persistence diagrams $D_1, D_2$ and the cost matrix
\begin{equation}
    C = 
    \begin{pmatrix}
        \hat{C} & u \\ v^T & 0
    \end{pmatrix} \in \mathbb{R}^{(n+1) \times (m+1)},
\end{equation}
where $\hat{C}$ is the pairwise distance matrix between points of $D_1$ and $D_2$, and $u$, $v$ are the distances of points in $D_1$, $D_2$ to the diagonal, respectively, Lacombe et al.~\cite{lacombe_al2018optimaltransport} show that the $2$-Wasserstein distance can be expressed by
\begin{equation}
    \label{Equation: Wasserstein distance as optimal transport}
    \wasserstein_2^2(D_1, D_2) = \textbf{L}_C(D_1 + R(D_2), D_2 + R(D_1)).
\end{equation}
The entropic regularized version of Eq.~\eqref{Equation: Wasserstein distance as optimal transport} gives rise to an approximation of the $2$-Wasserstein distance. A bound on the approximation error between $\wasserstein_2(D_1, D_2)$ and its regularized version is proven in~\cite{lacombe_al2018optimaltransport}.
\section{Random Sequential Adsorption}
\label{Appendix: RSA}

Our framework can be applied to synthetic data generated by an arbitrary method. In our experiments, we use the random sequential adsorption algorithm~\cite{chiu_al2013stochasticgeometry}, which is a standard method to generate systems of nonoverlapping particles. Given a set of particles, the algorithm sequentially inserts them into a bounded container. For each particle, uniformly generated candidate positions are tested iteratively. The candidate position is accepted if the particle does not intersect any previously placed particles. Otherwise a new candidate position is tested. If no admissible position is found after a fixed number of tries, the particle is rotated by permuting its axes and the placement procedure is repeated. If the particle cannot be placed in any rotation it is discarded and the algorithm proceeds to the next particle.

The set of particles can be generated by various models, e.g.\ as the typical cell of a fitted Gibbs--Laguerre tessellation~\cite{Schladitz_al2024particlemodelscoating} or as Gaussian random fields on the sphere~\cite{Feinhauer_al2015ParticleModellingSphericalHarmonics, Furat_al2021outershellmodel}. In this work, we refrain from modeling the particles and just rearrange the particles from the original data.

Checking different positions for valid particle placements scales with the number of voxels in a particle. Therefore, computational cost can be reduced drastically by reducing the image size. In order to both speed up computation and generate synthetic images with the full resolution, we generate two images, one in the original size and one with a downscaled size, when applying RSA. We generate a set of particles for the downscaled image by downsampling the original image with a majority vote in every downsample block. While applying the RSA algorithm, we test the positions in the downscaled image. If a downscaled particle is placed, we also place the original particle at the corresponding position in the full resolution image. 

While we always use the full resolution images in our experiments, maintaining the synthetic image at both resolutions has the additional benefit that the graph representation of the particle system can be computed on the downscaled resolution, followed by an upscaling of its features. This decreases the computational cost of the particle graph significantly.
\section{Experiment Details}
\label{Appendix: Experiment Details}

As described in Appendix~\ref{Appendix: RSA}, we compute pairs of a downscaled and full resolution synthetic image. The downscaled image size is obtained by scaling each spatial dimension by a factor of $4$. During particle placement, $12$ million random positions were tested for each particle and rotation. The computation is done on an NVIDIA A16 GPU.

The particle graphs were computed on the downscaled images using a threshold of $\tau = 100$, which corresponds to a threshold $\tau=400$ in the full resolution. Subsequently we scaled the node and edge features of the computed graphs by a factor of $4$ to recover the original spatial resolution. 

\begin{table}[t]
    \centering
    \textbf{Optimization Hyperparameters}
    \vspace{.2em}

    {\small
    \begin{tabular}{lc}
        \toprule
        Hyperparameter & Value \\
        \midrule
        Optimizer & Adam \\
        Learning rate & $5 \cdot 10^{-1}$ \\
        Learning rate scheduler & ExponentialLR \\
        Exponential decay factor & $0.99$ \\
        Number of epochs & $500$ \\
        Entropic regularization $\gamma$ & $100$ \\
        \bottomrule
    \end{tabular}
    }

    \caption{Hyperparameters used in the topological optimization during the experiments.}
    \label{Table: Hyperparameters Optimization}
\end{table}

The optimization framework is implemented using PyTorch~\cite{Paszke_al2019PytorchReference} for the gradient updates and GeomLoss~\cite{Feydy_al2019GeomlossPackage} for the loss term. The hyperparameters used during the optimization are listed in Table~\ref{Table: Hyperparameters Optimization}. The computations are done entirely on an Intel(R) Xeon(R) Gold 6354 CPU.

\end{document}